\documentclass{aa} % for a referee version
\usepackage{diagbox}
\usepackage{array}
\usepackage{booktabs}
\usepackage{graphicx}
\usepackage{multirow}
\usepackage{xcolor}
\usepackage{soul}
\usepackage{bm}
\usepackage[justification=centering]{caption} 
\usepackage{caption}
\usepackage{subcaption}

\usepackage{float}
\usepackage{stfloats}
\usepackage{graphicx}
\usepackage{natbib}

\usepackage{txfonts}
\usepackage{braket}
\def \be {\begin{equation}}
\def \ee {\end{equation}}
\def\restriction#1#2{\mathchoice
              {\setbox1\hbox{${\displaystyle #1}_{\scriptstyle #2}$}
              \restrictionaux{#1}{#2}}
              {\setbox1\hbox{${\textstyle #1}_{\scriptstyle #2}$}
              \restrictionaux{#1}{#2}}
              {\setbox1\hbox{${\scriptstyle #1}_{\scriptscriptstyle #2}$}
              \restrictionaux{#1}{#2}}
              {\setbox1\hbox{${\scriptscriptstyle #1}_{\scriptscriptstyle #2}$}
              \restrictionaux{#1}{#2}}}
\def\restrictionaux#1#2{{#1\,\smash{\vrule width 0.8pt height 1.\ht1 depth 1.\dp1}}_{\,#2}}

\newcommand{\genec}{\textsc{\large genec\normalsize}}

\newcommand{\mesaa}{\textsc{\large mesa\normalsize}}

\newcommand{\Genec}{\textsc{\large genec \normalsize}}

\defcitealias{pol98}{P98}

\defcitealias{pol98}{P98}

\defcitealias{hur00}{H00}

\defcitealias{hur00}{H00}

\defcitealias{hur02}{H02}

\defcitealias{hur02}{H02}

\defcitealias{sci24}{S24}

\defcitealias{sci24}{S24}

\defcitealias{zah77}{Z77}

\defcitealias{zah77}{Z77}

\usepackage[colorlinks=true, citecolor=blue, linkcolor=blue, urlcolor=blue]{hyperref}

\begin{document}
\title{Chemical evolution of close massive binaries - tidally-enhanced or tidally-suppressed mixing?}

\author{Luca Sciarini\inst{1}, Sophie Rosu\inst{1},
Sylvia Ekstr\"om\inst{1}, Maxime Marchand\inst{1}, Patrick Eggenberger\inst{1}, Georges Meynet\inst{1}}
\institute{Department of Astronomy, University of Geneva, Chemin Pegasi 51, CH-1290 Versoix, Switzerland\\
              \email{luca.sciarini@unige.ch}
         }

   \date{Received date ... /
Accepted date ...}
\authorrunning{Luca Sciarini et al.}
\titlerunning{Chemical evolution of close massive binaries - tidally-enhanced or tidally-suppressed mixing?}

  \abstract
   {One of the largest source of uncertainties in the predictions of stellar models comes from the internal transport mechanisms. In close massive binaries, previous theoretical studies suggest that tides consistently boost chemical mixing. However, observations do not reveal any clear period-nitrogen enrichment trend, challenging these predictions. In addition, comprehensive examinations of the interplay between tidal interactions, angular momentum and chemicals transport have so far been very scarce.}
   {Our goal is to investigate the interplay between tidal interactions and rotational mixing, and the impact of the angular moment transport (AMT) assumptions. We also aim to tackle the question of whether tidal interactions enhance or suppress chemical mixing.} 
   {We compute grids of \Genec binary models  with various AMT treatments at solar metallicity. In order to independently assess the role of tidal interactions, we systematically compute model variations of single stars with identical initial conditions.}
   {Our investigations reveal that tidal interactions can either enhance or suppress mixing relative to single-star models with identical initial conditions, and that the outcome is highly sensitive to the adopted AMT assumptions. We identify a key contrast between the two types of computed models: in close systems subject to tides, magnetic models predict that the mixing efficiency is mostly determined by the orbital configuration, whereas in hydrodynamic models it also depends on the assumed initial velocity. As a result, hydro models may display non-monotonic period–enrichment trends, or even period-enrichment correlations.}
   {These results highlight the importance of the AMT assumptions in modeling binaries with tidal interactions, notably in the context of the chemically homogeneous evolution channel. The sensitivity of the predictions of hydro models to initial conditions extends the size of the period-enrichment parameter space they cover, allowing them to accommodate for peculiar observed systems, i.e., with mild enrichment at short periods, or high enrichment at longer periods.}

   \keywords{
                stars: abundances --
                stars: evolution --
                stars: massive --
                stars: rotation -- (stars:) binaries: general --
                (stars:) binaries (including multiple): close
               }

   \maketitle

%________________________________________________________________

\section{Introduction}
Stellar models are crucial in the interpretation and analysis of astrophysical data \citep[e.g.,][]{bro11a,bre12,eks12,cho16,lim18,ama19,cos25}. One of the largest sources of uncertainties in their predictions comes from the internal transport mechanisms \citep[e.g.,][]{bul19,ful19,egg24}. Asteroseismology has proven successful in constraining the AMT efficiency in low to intermediate-mass stars \citep[e.g.,][]{deh14,van16,pap17,aer19}. These observations generally indicate that an efficient AMT is at play in radiative zones, which cannot be reproduced solely by hydrodynamic instabilities \citep[e.g.,][]{egg12,egg22,ful19,moy23}. In contrast, early-B and O-stars seem to show a diversity of internal rotation profiles, with asteroseismic observations of $\beta$-Cephei stars revealing both radial differential rotation and uniform rotation \citep[see e.g.,][]{aer03,dup04,bri07,dzi08,mar13,oua20,sal22,bur23,fri25}.

It has long been identified that main-sequence (MS) massive stars often show observational evidences of carbon-nitrogen-oxygen (CNO) cycle signatures, suggesting that efficient mixing operates in their radiative envelope \citep[e.g.,][]{gie92,kil92,mor08,hun09,bro11b,gri17,mah20}. Various physical processes have been proposed to explain these signatures, notably rotational mixing \citep[RM, e.g.,][]{mae00,bro11b}, internal gravity waves \citep[IGWs, e.g.,][]{rog17,mom25} and binary interactions \citep[e.g.,][]{dem09,lan12}.

The majority of massive stars possess companions; many of these systems are expected to interact at some stage of their evolution \citep[e.g.,][]{san12,moe17,off23,bod25,pat25}. \citet{dem09} proposed to use pre-interacting close binary systems to test RM. In these close configurations, tides are expected to synchronize the stars to the orbital motion, which should boost RM. This theoretical prediction was confirmed by the studies of \citet{son13,son16}, who found that tidal interactions always boost RM. However, observational studies challenge this picture: the majority of the short-period systems in the samples of \citet{mar17,pav18,pav23,abd19,abd21} do not show any evidence of strong nitrogen enrichment. Moreover, they do not suggest a clear anti-correlation between period and nitrogen enrichment --a direct prediction of the models by \cite{dem09}.

In this study, we investigate the interplay between tidal torques and RM, with a particular focus on the impact of the AMT assumptions. To this aim, we compute several single and binary star grids with the GENeva Evolutionary Code \citep[\genec,][]{egg08} with different AMT treatments: purely-hydrodynamic models with an advective-diffusive AMT and magneto-hydrodynamic models accounting for the calibrated Tayler-Spruit instabilities \citep{egg22}. Tides are often seen as an additional, independent source of chemical mixing in radiative envelopes However, in the way they are accounted for in most state-of-the-art binary codes (e.g., in \mesaa, \citealt{pax15,pax18,pax19,mar16}, \textsc{\large bonn, \normalsize}\citealt{dem09,yoo10,sze22}, or \genec,  \citealt{son13,son16}), they only alter mixing by modifying the stellar rotation, and thereby the RM efficiency (but this picture may be too simplistic, see discussion in Sect. \ref{caveats}). We aim to demonstrate that in this framework, model reactions to tidal torques strongly depend on the AMT treatments and that in certain configurations, stars in binary systems may experience less mixing --"tidally-suppressed mixing"-- than single stars with identical initial conditions. The approach and conclusions of the present paper are reminiscent of \citet{zah94}, who demonstrated that tidal synchronization in late-type binaries reduces RM, thereby providing a theoretical explanation for their reduced lithium depletion relative to single-star counterparts.

The structure of this paper is as follows: in Sect. \ref{methods}, we provide a description of our methods, and the single and binary star physics assumptions we made. In Sect. \ref{results}, we present our binary models with tidal mixing. In Sect. \ref{discussion}, we discuss the limitations of our results and how they compare to those of previous studies and observations. Finally, we summarize our findings in Sect. \ref{conclusion}.
\section{Single and binary star physics}\label{methods}
\subsection{Stellar physics ingredients}\label{ingredients}
In this study we use \Genec to compute detailed stellar evolution 1D models. The adopted stellar physics is generally the same as in \citet{eks12}, although with some differences that are explicitly listed below. We recall here the main ingredients.

The adopted wind prescription is the standard combination used in the \Genec grids. It consists of the \citet{dej88}, \citet{vin01} and \citet{nug00} prescriptions lowered by a scaling factor 0.85.

During the MS, convection in the core is assumed to be very efficient, i.e., angular momentum transport and chemical mixing are assumed to be instantaneous in this region. Small convective zones near the stellar surface are in contrast treated following the mixing length theory (MLT) framework, with $\alpha_{\rm MLT}=1.6$. The size of the convective core is increased following a step overshoot treatment with $\alpha_{\rm ov}=0.1$.

We computed models with two different AMT treatments. Purely-hydrodynamic models\footnote{\Genec is a hydrostatic code; however, the secular hydrodynamical instabilities induced by rotation are taken into account by solving an advective–diffusive equation for the AMT. We refer to these as "hydrodynamic models".} (hereafter hydro models) account for shear instabilities \citep{zah92} and meridional circulation \citep[][]{edd25,swe50,zah92,mae98}. Magneto-hydrodynamic models (hereafter magnetic models) in addition account for the Tayler-Spruit instability \citep{tay73,spr02,mae03,mae04,mae05}. For the magnetic models we used the asteroseismic-calibrated version of the Tayler-Spruit dynamo as formulated by \citet{egg22}.

In hydro models, the AMT by meridional circulation is treated as an advective process as in \citet{eks12}. In this case, the advective-diffusive equation reads:
\begin{equation}
    \rho\frac{\partial\left(r^2\Omega\right)}{\partial t}=\frac{1}{5r^2}\frac{\partial}{\partial r}\left(\rho r^4\Omega U(r)\right)+\frac{1}{r^2}\frac{\partial}{\partial r}\left(\rho r^4\Omega D_{\rm AM}\frac{\partial \Omega}{\partial r}\right),
\end{equation}
where $\rho$ is the mean isobar density, $r$ the radius, $U$ the radial dependence of the vertical component of the meridional circulation, $\Omega$ the average angular velocity on an isobar, $D_{\rm AM}$ the diffusion coefficient for AMT. As demonstrated in \citet{mae05}, when the magnetic instabilities are accounted for, they are strong enough for achieving near solid-body rotation during the whole MS evolution, so that the AMT by meridional circulation can effectively be neglected and a fully diffusive treatment can be followed.

\Genec models also account for the RM imparted by the three mentioned processes. In the present work, shear instabilities were modeled using the expression of \citet{mae97}:
\begin{equation}
    D_{\rm shear}=f_{\rm energ}\frac{H_P}{g\delta}\frac{K}{\frac{\phi}{\delta}\nabla_\mu + \nabla_{\rm ad}-\nabla_{\rm rad}}\left(\frac{9\pi}{32}\Omega\frac{\text{d}\ln\Omega}{\text{d}\ln r}\right)^2,
    \label{déchire}
\end{equation}
where $f_{\rm energ}$ is the fraction of the excess energy in the shear that contributes to mixing, $H_{P}$ the pressure scale height, $g$ the local gravity, $\delta$ and  $\phi$ thermodynamic derivatives, $K$ the thermal diffusivity, $\nabla_\mu$, $\nabla_{\rm ad}$ and $\nabla_{\rm rad}$ respectively the mean molecular weight, adiabatic and radiative gradients. It is worth noting that the shear diffusion coefficient depends on both $\Omega$ and its gradient.

As demonstrated by \citet{cha92}, the combined effect of the meridional circulation and the horizontal turbulence mitigates the advection of the chemical elements through homogenization of horizontal layers. The vertical RM then reduces to a diffusion process, which can be described by the effective diffusion coefficient:
\begin{equation}
    D_{\rm eff}=\frac{1}{30}\frac{\left|rU(r)\right|^2}{D_{\rm h}}.
    \label{déaiffe}
\end{equation}
The horizontal turbulence was modeled using the prescription of \citet{zah92}:
\begin{equation}
    D_{\rm h}=\frac{1}{c_{\rm h}} r\left|2V(r)-\alpha U(r)\right|,
    \label{déhâche}
\end{equation}
where $c_{\rm h}$ is a constant of the order of 1, $V$ the radial dependence of the horizontal component of the meridional circulation and $\alpha = \frac{1}{2}\frac{\text{d}\ln(r^2\bar\Omega)}{\text{d}\ln r}$. 
The transport equation for the chemicals therefore reads:
\begin{equation}
\rho\frac{\partial X_i}{\partial t}=\frac{1}{r^2}\frac{\partial}{\partial r}\left(\rho r^2 D_{\rm chem}\frac{\partial X_i}{\partial r}\right),
\end{equation}
with $X_i$ the considered chemical species, $D_{\rm chem}=D_{\rm AM}+D_{\rm eff}$. We note that the diffusive treatment only holds for RM. In magnetic models where the AMT is highly efficient, the contribution of meridional circulation to the AMT can be neglected; however, its impact on RM remains significant. The shear diffusion coefficient is negligible for flat $\Omega$-profiles (see Eq. \eqref{déchire}), as a result RM is in fact dominated by meridional circulation, i.e., $D_{\rm chem}\approx D_{\rm eff}$ \citep[][see also Fig \ref{profiles}]{mae05,son16,nan24,asa25}. In this case, the RM efficiency is primarily determined by the angular momentum (AM) content of the stars (or equivalently their angular velocity).

Following this treatment of AMT and RM, the presented hydro models are computed with the same physics as in \citet{eks12}. These models were calibrated to reproduce the observed abundances of nitrogen of a population of Galactic MS B-type stars. We performed an equivalent calibration for the magnetic models. We found that using $c_{\rm h}=0.70$ in Eq. \eqref{déhâche} (instead of $c_{\rm h}=1$ for the hydro models) offers a satisfactory reproduction of the observed abundances of the same stars. The details of the performed calibration are given in Appendix \ref{AppA}. As a result of this calibration, the predicted RM of magnetic models is reduced compared to those presented in \citet{son16}, which should alter the size of the predicted parameter space where chemically homogeneous evolution occurs in close binaries.
\subsection{Binary interactions}\label{binary_interactions}
We simulate binary systems in circular orbits. We effectively evolve single stars but account for binary interactions assuming a non-evolving point mass for the companion, as in e.g., \citet{det08,dem09,son13,son16,son18,sci24}. The simulations are stopped either when the star fills its Roche lobe (RL), which size is computed following \citet{egg83}, or at terminal-age-main-sequence (TAMS) if the RL has not been filled during the MS evolution.

We propose a refined treatment of tides in massive \Genec models, following the recent findings of \citet{fra23,sci24}. The dynamical tides are treated as in \citet{son13,sci24}, following the \citet{zah77} prescription. In case of circular orbits, the tidal torque is:
\begin{equation}
\restriction{\frac{\text{d}}{\text{d}t}\left(I\Omega_{\rm spin}\right)}{\rm Dyn}=\frac{3}{2}\frac{GM^2}{R}E_2\left(q^2\left(\frac{R}{a}\right)^6\right)s_{22}^{8/3}\text{sgn}(s_{22}),
\label{correct}
\end{equation}
where $I$ is the moment of inertia of the star, $\Omega_{\rm spin}$, $M$ and $R$ respectively its surface angular velocity, mass, and radius, $G$ the gravitational constant, $q$ the mass ratio, $a$ the semi-major axis of the orbit. For $E_2$, we use the prescription by \citet{qin18}. Following the approach of \citet{son13,son16,son18}, the tidal torques are applied to an outer layer comprising 3\% of the total mass, assumed to rotate rigidly. This simplification provides a numerically stable way to deposit the torque in the envelope while allowing differential rotation in the interior. In reality, the dynamical tides torque results from gravity waves traveling trough the envelope \citep{zah75,zah77, gol89} and dissipating either gradually or when reaching critical layers \cite[e.g.][]{tal98,mat09}. Our treatment should thus be viewed as an approximation of AM deposition in the outer radiative envelope, rather than a detailed physical model of wave damping.

It is worth highlighting that $s_{22}$ scales as $\Omega_{\rm orb}-\Omega_{\rm spin}$, with $\Omega_{\rm orb}$ and $\Omega_{\rm spin}$ the orbital and spin angular velocities. Hence, the dynamical tides torque scales as $(\Omega_{\rm orb}-\Omega_{\rm spin})^{8/3}$. As demonstrated in \citet{sci24}, in the \citet{hur02} adaptation of the \citet{zah77} formalism the dynamical tides torque in contrast scales linearly with $\Omega_{\rm orb}-\Omega_{\rm spin}$, which alters the torque efficiency depending on the departure from synchronization. When the dynamical tides are treated consistently with the original formulation by \citet{zah77}, they are very efficient when the star is far from synchronization, but become extremely inefficient close to synchronization \citep[see Figs. A.1 and A.2 in][]{sci24}. As a result, under this formalism the stars may deviate from synchronization at Roche lobe overflow (RLOF).

\citet{fra23} proposed that equilibrium tides acting on small convective zones near the surface can become the dominant contribution to the evolution towards synchronization already during the MS due the fast decrease of the term $E_2$. In the present work we modeled equilibrium tides acting on small sub-surface convective zones following a treatment similar to that in \citet{fra23}. The treatment of equilibrium tides in \citet{fra23} is inspired from \citet{hur02}, which is adapted from \citet{hut81,ras96}. Under their formalism, the ratio of the apsidal motion constant (AMC) to the tidal dissipation timescale $k_2/T$ in the \citep{hut81} formulas is replaced by:
\begin{equation}
    \left(\frac{k_2}{T}\right)_{\rm conv}=\frac{2}{21}\frac{f_{\rm conv}}{\tau_{\rm conv}}\frac{M_{\rm conv.reg.}}{M}, \ \ f_{\rm conv}=\min\left(1,\left(\frac{P_{\rm tid}}{2\tau_{\rm conv}}\right)^2\right),
   \label{posydon}
\end{equation}
where $\tau_{\rm conv}$ is the convective turnover timescale, $M_{\rm conv.reg.}$ the mass of the convective region, $M$ the total mass of the star. $f_{\rm conv}$ reduces the strength of the tides when the turnover timescale is greater than the pumping timescale $P_{\rm tid}=2\pi\left|\Omega_{\rm orb}-\Omega_{\rm spin}\right|^{-1}$ \citep[see e.g.,][]{zah66,gol77,goo97,zah08,sou13}. Eq. \eqref{posydon} is obtained comparing the circularization timescales of \citet{hut81} and \citet{ras96}. By doing so, the AMC dependence is lost.

In this work, we implemented a formalism closer to the original formulation by \citet{hut81}, keeping the AMC dependence. We include the term $f_{\rm conv}$ in our formalism to account for fast tides. Furthermore, as tides involve AM exchanges, we scale the strength of the tides by the ratio of the moment of inertia of the convective region $I_{\rm conv.reg}$ to that of the star $I$ instead of the mass ratio. As a result, our equilibrium tides prescriptions for small sub-surface convective regions can be written as:
\begin{equation}
    \left(\frac{k_2}{T}\right)_{\rm conv}\rightarrow \frac{k_2}{\tau_{\rm conv}}f_{\rm conv}\frac{I_{\rm conv.reg.}}{I}.
    \label{prescription}
\end{equation}
The term 2/21 is not present in our prescription as it is not present in the original formalism by \citet{hut81}. The tidal torque in case of circular orbits is thus:
\begin{equation}
    \restriction{\frac{\text{d}}{\text{d}t}\left(I\Omega_{\rm spin}\right)}{\rm Eq}=
    3\frac{k_2}{\tau_{\rm conv}}f_{\rm conv}\frac{I_{\rm conv.reg.}}{I}MR^2q^2\left(\frac{R}{a}\right)^6 \left(\Omega_{\rm orb}-\Omega_{\rm spin}\right).
\label{eq_tides}
\end{equation}
We compute $k_2$ consistently with the stellar structure solving the Clairaut-Radau equation:
\begin{equation}
    \begin{split}
    r\frac{\text{d}\eta_2(r)}{\text{d}r}&=\eta_2(r)-\eta_2^2(r)+6\left(1-\frac{\rho(r)}{\bar\rho(r)}\left(\eta_2(r)+1\right)\right),\\
    k_2&=\frac{3-\eta_2(R)}{4+2\eta_2(R)},
    \end{split}
    \label{cla-rad}
\end{equation}
where $\rho$ and $\bar\rho$ are respectively the density and average density at distance $r$ from the center and $R$ is the radius at optical depth $\tau~=~ 2/3$. Eq. \eqref{cla-rad} is solved using a 4-th order Runge-Kutta integrator. The AMC is directly linked to the internal density profiles of the stars and serves as a powerful observable constraint, being related to the apsidal motion of close binary systems \citep[e.g.,][]{ste39,cla04,cla24,cla10,ros20a,ros20b,ros22a,ros22b}. This quantity is included among the outputs of the stellar grids provided with this paper (Sect. \ref{results}). The tracks are available on \href{https://zenodo.org/records/18302392}{Zenodo}, \href{https://doi.org/10.26037/yareta:gejbckay45bjhnrzklcev6tb6u}{Yareta}, and on the \href{https://www.unige.ch/sciences/astro/evolution/en/database}{Geneva stellar group database}.

It is worth noting that the \citet{hut81} formalism relies on the simplifying assumption of constant time lag. As demonstrated by \citet{zah08}; \citet{rem12}, this approximation breaks down in the fast tides regime ($\tau_{\rm conv}>P_{\rm tid}$). In this case it is in principle needed to expand the tidal potential in its multiple Fourier components. However, as illustrated in Appendix \ref{AppB}, we find that the dynamical (equilibrium) tides largely dominate when the stars are far from (close to) synchronization. As such, $\tau_{\rm conv}<P_{\rm tid}$ is always verified when equilibrium tides dominate.

Finally, in \citet{hur02,fra23}, the convective turnover timescale is obtained using an estimate adapted from \citet{ras96}. In \citet{fra23}, they used:
\begin{equation}
    \scalebox{1.1}{$
    \tau_{\rm conv}=\left(\frac{M_{\rm conv.reg.}}{3L}\frac{R_{\rm t,conv.reg.} + R_{\rm b,conv.reg.}}{2}\left(R_{\rm t,conv.reg.}-R_{\rm b,conv.reg.}\right)\right)^{1/3}.
    \label{pos}$}
\end{equation}
We note that this timescale directly depends on the mass $M_{\rm conv.reg.}$ and radial extent $R_{\rm t,conv.reg.}-R_{\rm b,conv.reg.}$of the convective region. As such, it can get very small depending on the size of the considered convective zone. In this work, we obtained the value of $\tau_{\rm conv}$ within the MLT framework (see Sect. \ref{ingredients}) as:
\begin{equation}
    \tau_{\rm conv}=\frac{l_{\rm conv}}{\upsilon_{\rm conv}},
\end{equation}
where $l_{\rm conv}=\alpha_{\rm MLT}H_{\rm P}$ and $\upsilon_{\rm conv}$ are the local mixing length and velocity of the bubble in the MLT framework. In a convective region where the local value of $\tau_{{\rm conv}}$ differs among several shells, we take the minimum value across the convective region. Models presented in this work typically have several sub-surface convective regions. We account for the combined effect of all the regions by adding up the tidal torques. The evolution toward synchronization of an illustrative system --considering only dynamical tides or both dynamical and equilibrium tides, treated either following \citet{fra23} or Eq. \eqref{prescription}-- is shown in Appendix \ref{AppB}.

We account for orbital evolution imparted by mass-loss and spin-orbit AM exchange according to:
\begin{equation}
    \frac{\dot a}{a}=2\frac{\dot J_{\rm orb}}{J_{\rm orb}}-2\frac{\dot M_1}{M_1}+\frac{\dot M_1}{M_1+M_2},
    \label{dota}
\end{equation}
where $M_1$ and $M_2$ are the masses of the stars, $a$ the semi-major axis of the orbit and $J_{\rm orb}$ the orbital AM. Eq. \eqref{dota} is obtained following an AM balance approach under the assumption that $\dot M_2=0$. $\dot J_{\rm orb}$ is expressed as:
\begin{equation}
\begin{split}
    &\dot J_{\rm orb}=\dot J_{\rm tides}+\dot J_{\rm ml},\\
    &\text{where }\dot J_{\rm tides} = -\left(\restriction{\frac{\text{d}}{\text{d}t}\left(I\Omega_{\rm spin}\right)}{\rm Dyn}+\restriction{\frac{\text{d}}{\text{d}t}\left(I\Omega_{\rm spin}\right)}{\rm Eq}\right)
    \end{split}
\end{equation}
and $\dot J_{\rm ml}$ is the AM carried away by mass loss assuming that the removed mass has the specific orbital AM of the star.

In case of twin (equal mass) systems, we assume the companion has the same mass loss and initial parameters so that its contribution to the orbital evolution is the same as that of the simulated star. In this case, the right hand side of Eq. \eqref{dota} is simply multiplied by two.
\section{Binary models with tidal mixing}\label{results}
We present several sets of binary models at solar metallicity ($Z_\odot=0.014$), incorporating the ingredients described in Sect. \ref{methods}, to illustrate how tides can either enhance or suppress RM, and how closely this behavior is linked to the choice of AMT mechanism. To this end, we systematically compare the outputs of magnetic and hydro models, using single-star models as a reference.
\subsection{Comparison with \citet{son13}}\label{comp_with_song}
We start by reproducing the results of \citet{son13}\footnote{The hydro models are not exactly identical to those of \citet{son13} as the equilibrium tides were not accounted for in their study.} and comparing them with magnetic models. This allows us to illustrate in a simple way that magnetic and hydro models react differently to tidal torques. We provide illustrative examples of the evolution toward synchronization for both types of models in Appendix \ref{AppC}.
\subsubsection{Spin-down models}\label{spin_down}
\begin{figure*}[h]
\centering
\centerline{\includegraphics[trim=0cm -0.2cm 0cm 0cm, clip=true, width=2.0\columnwidth,angle=0]{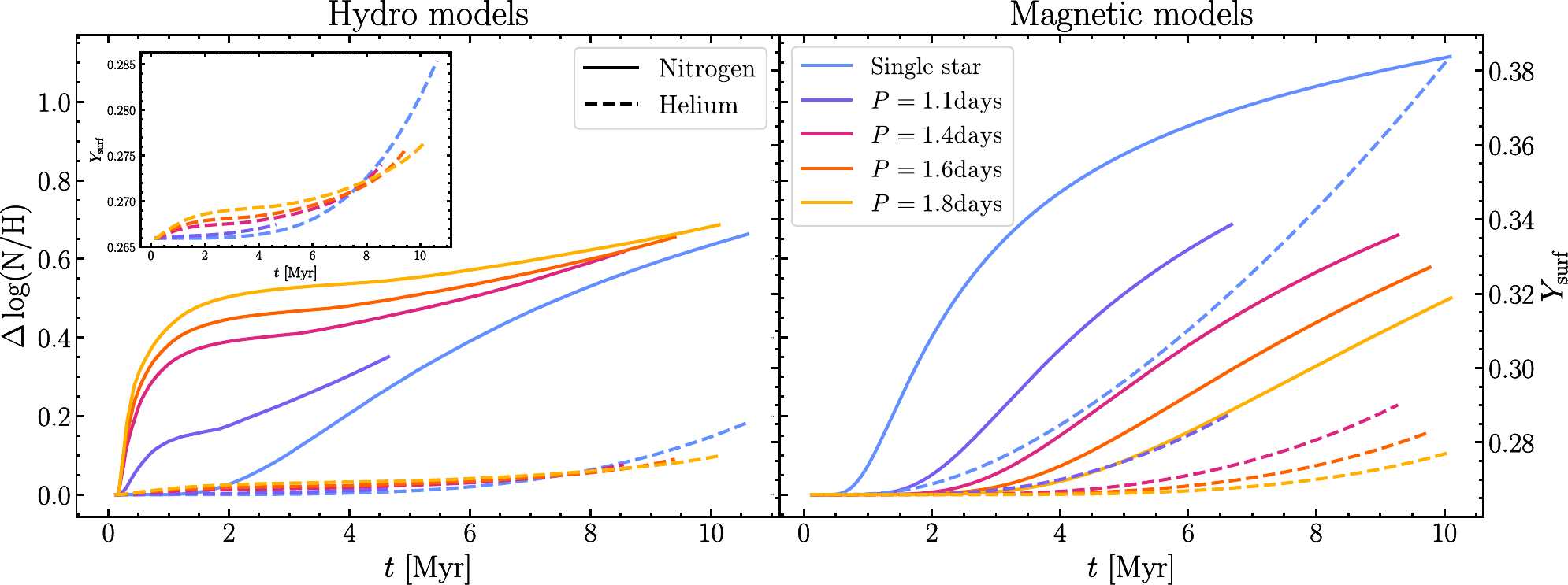}}
\caption{Nitrogen and helium enhancement evolution of the same spin-down systems as in \citet{son13}.
\textit{Left panel:} hydro models. \textit{Right panel:} magnetic models.}
\label{son_spin_down}
\end{figure*}

\begin{figure}[h]
\centering
\centerline{\includegraphics[trim=2.1cm 2.6cm 1.8cm 2.2cm, clip=true, width=1.0\columnwidth,angle=0]{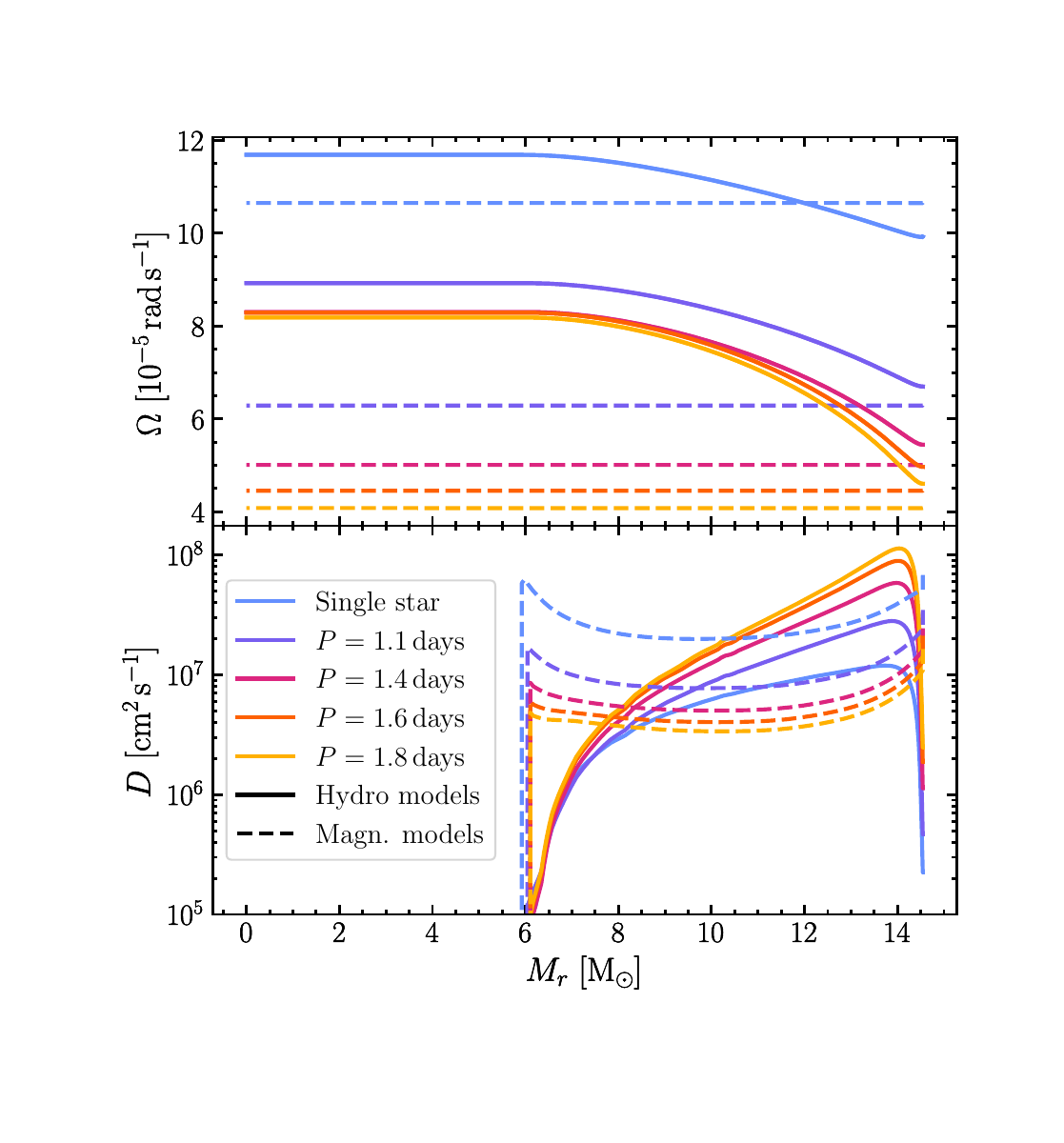}}
\caption{\textit{Upper panel:} $\Omega$ profiles of the hydro and magnetic models at $t=500$\,kyrs of the same spin-down systems as in \citet{son13}.
\textit{Lower panel:} diffusion coefficient profiles.}
\label{profiles}
\end{figure}

The spin down-models presented in \citet{son13} are primaries of initial mass $M_1=15$\,M$_\odot$, with a companion of fixed mass $M_2=10$\,M$_\odot$. The $15$\,M$_\odot$ models are initiated at 60\% of the critical velocity, defined as in \citet{eks12}. The chosen periods are $P=1.1,1.4,1.6$ and 1.8\,days. Fig. \ref{son_spin_down} shows the nitrogen enhancement, defined as:
\begin{equation}
    \Delta\log\left(\text{N}/\text{H}\right)=\left[\text{N}/\text{H}\right]-[\text{N}/\text{H}]_0,
\end{equation}
and the surface helium mass fraction $Y_{\rm surf}$ evolution. Both the nitrogen and the helium enhancement are used as tracers of RM.
Fig. \ref{son_spin_down} allows to observe several key results:
\begin{enumerate}
    \item In this spin-down configuration, tides increase mixing in hydro models and decrease mixing in magnetic models. Compared to single-star evolution, hydro models accounting for spin-down by tides are more enhanced in both nitrogen and helium (the picture is more complex for helium, see remark 4). In contrast, magnetic binary models show less enrichment in both nitrogen and helium than the single-star model. 
    \item The nitrogen enrichment of hydro models increases with increasing period, which can seem counterintuitive as a longer period corresponds to a lower synchronization angular velocity. Mixing is more efficient at lower synchronization angular velocities because they generate a higher degree of differential rotation ($\Omega$-gradient), which increases shear mixing (see Eq. \eqref{déchire} and Fig. \ref{profiles}).
    \item The nitrogen and helium enrichments of magnetic models decrease with increasing period, which can be explained by the fact that the strong AMT in magnetic models makes shear mixing inefficient. The $\Omega$-profiles are flat, and hence mixing is not increased by the tidal torques, but decreases as the stars spin down (see Fig. \ref{profiles}).
    \item The nitrogen and helium enhancement of hydro models do not follow exactly the same evolution. Nitrogen is efficiently enhanced at the beginning of the evolution of models with tides, while the simultaneous helium enrichment is less pronounced. Hydro models are subject to high degrees of differential rotation at the beginning of the evolution caused by the tidal torques, which lead to strong shear mixing. Nitrogen is more sensitive to this early shear as the nitrogen profile quickly changes due to CNO burning. Although helium is also impacted by CNO, the profiles are less steep than the nitrogen profiles and thus RM at the beginning of the evolution is less efficient in enhancing He. The single-star model ends up more enriched in He than the binary models with $P=1.4,1.6$ and 1.8\,days as it has kept more AM and therefore He is more efficiently mixed later in the evolution, when the He profile becomes steeper (see also Sect. \ref{song_massive} and Appendix \ref{AppD}).
\end{enumerate}
Fig. \ref{profiles} shows the angular velocity and diffusion coefficient profiles at $t=500$\,kyrs of the hydro and magnetic models (the shear diffusion coefficient for the former, the effective diffusion coefficient for the latter).
It clearly illustrates the points discussed above: magnetic models have flat rotation profiles and therefore mixing is dominated by meridional circulation ($D_{\rm eff})$. Even when subject to strong tidal torques, their profiles remain flat and therefore they are not subject to shear mixing. Systems of shorter periods are more braked and therefore have lower diffusion coefficient profiles. In hydro models, the AMT is less efficient and the stars rotate differentially. The degree of differential rotation increases with increasing period, and as a result $D_{\rm shear}$ is larger for longer periods. 
\subsubsection{Spin-up models}
\begin{figure*}[h]
\centering
\centerline{\includegraphics[trim=5.1cm 1.2cm 1.4cm 0.8cm, clip=true, width=1.65\columnwidth,angle=0]{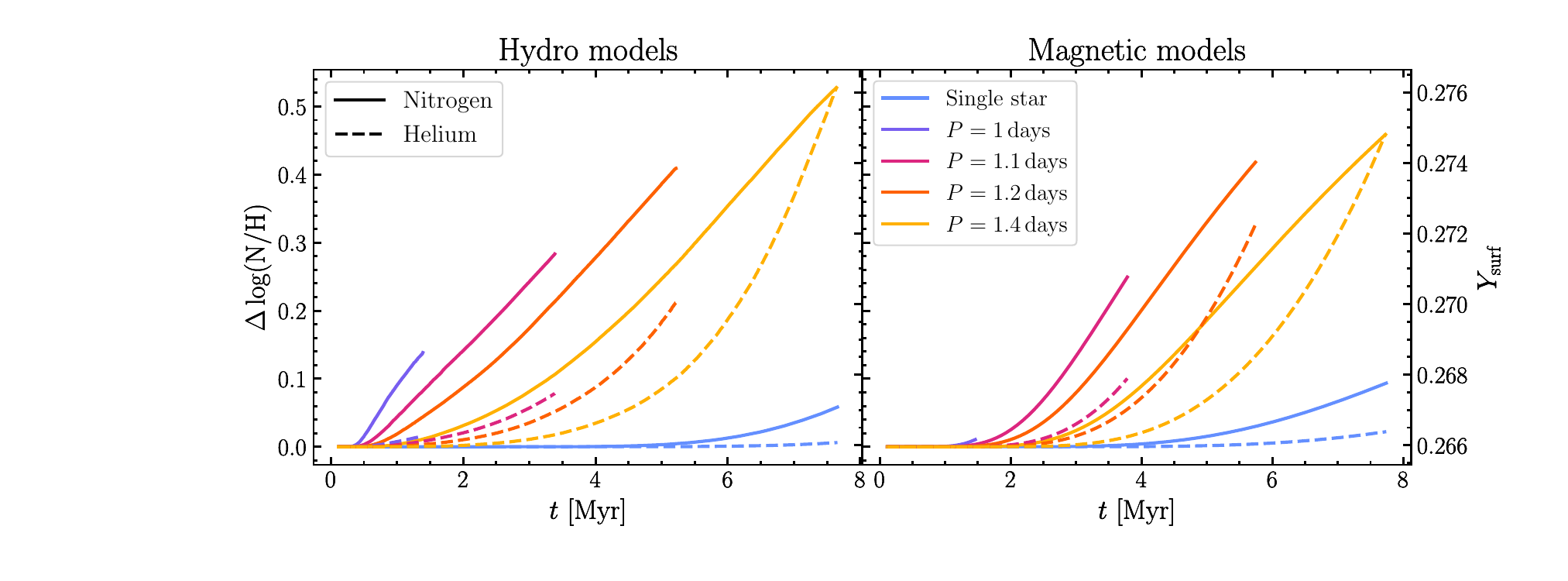}}
\caption{Same as Fig. \ref{son_spin_down} for spin-up models.}
\label{son_spin_up}
\end{figure*}

We here focus on spin-up models, using the same initial conditions as in \citet{son13}, and compare the results with magnetic models. In this case, the initial velocity is $\upsilon_{\rm ini}/\upsilon_{\rm crit}=0.2$, the periods $P=1,1.1,1.2$ and 1.4\,days, and the initial masses are the same as in Sect. \ref{spin_down}. Fig. \ref{son_spin_up} shows the nitrogen and helium enhancements evolution.
In this case hydro and magnetic models both predict that tides increase mixing. This illustrates that the impact of tidal torques on RM is simpler in magnetic models: in spin-up cases, they enhance mixing, whereas in spin-down cases, they suppress mixing. The $\Omega$-profile of magnetic models being flat, RM by shear is inefficient and mixing only depends on the value of $\Omega$. 

In the case of hydro models, the picture is more complex: the efficiency of RM depends on both the value and the gradient of $\Omega$. Spin-up by tides momentarily decreases the differential rotation, but it also increases the value of $\Omega$ (see Fig. 12 in \citealt{son13}). The net effect in this case is an increase in mixing with increasing synchronization angular velocity (decreasing period). We show in Sect. \ref{grid} that with hydro models there also exists configurations where spun-up models are less enriched than their single-star counterparts, a result that may appear even more counterintuitive.
\subsubsection{Spin-down massive star models}\label{song_massive}
\begin{figure*}[h]
\centering
\centerline{\includegraphics[trim=5.1cm 1.2cm 1.4cm 0.8cm, clip=true, width=1.65\columnwidth,angle=0]{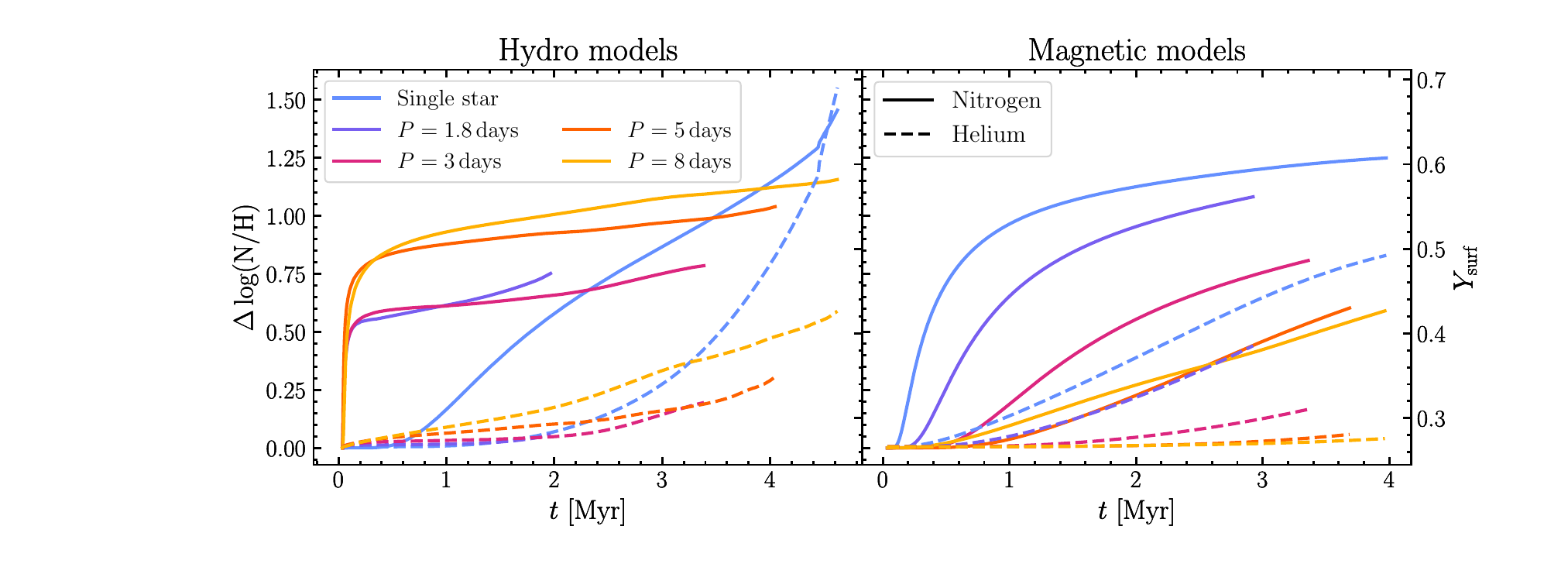}}
\caption{Same as Fig. \ref{son_spin_down} for models with $M_1=45$\,M$_\odot$, $M_2=30$\,M$_\odot$ and $P=1.8,3,5$ and 8\,days.}
\label{spin_down_massive}
\end{figure*}

In order to test the effect of mass on the previous results, we compute spin-down models with the same initial velocity as the models of Sect. \ref{spin_down}, but with larger initial masses ($M_1=45$\,M$_\odot$ and $M_2=30$\,M$_\odot$) and longer periods ($P=1.8,3,5$ and 8\,days). Fig. \ref{spin_down_massive} shows the nitrogen and helium enhancements evolution of the hydro and magnetic models.
We note that although the same trends as in Fig. \ref{son_spin_down} are observed, the conclusions are different. The magnetic models still behave the same: spun-down stars are less enriched than the single star, and the longer the period (lower $\Omega_{\rm orb}$) the lower the enrichment. 

Hydro models also show the same trend as in Sect. \ref{spin_down} (the longer the period, the higher the enrichment). However, while tidal spin-down initially boosts shear mixing --mostly enhancing nitrogen-- after this phase of fast enrichment the nitrogen enhancement slows. At RLOF binary models are less enriched in both helium and nitrogen than the single-star model, for which the enrichment steadily increases throughout the evolution. This can be attributed to the fact that in the long term, mixing is quenched in the binary models as the tidal torques make them rotate at lower angular velocity and reduce the $\Omega$--gradients (once the stars are synchronized, the tidal torque acts in the opposite direction, transferring AM from the orbit to the spin in order to maintain synchronization. By keeping the surface synchronized, this prevents the development of significant differential rotation).

The difference in enrichment between the single star and the binary systems is more pronounced for helium. For the single star, the helium enrichment accelerates at the end of the evolution, as the He profile gets steeper. Models including tides synchronize quickly and as a result show less helium enrichment, as the angular velocity and the differential rotation have significantly decreased when CNO burning establishes a gradient in the helium profile (see Fig. \ref{profiles_massive} in Appendix \ref{AppD} for more details).

These results highlight the complexity of the reaction of hydro models to tidal torques. Depending on the configuration, spin-down systems may experience more or less mixing than the single star with same initial conditions.
\subsection{Grids of binary systems with tidal interactions}\label{grid}
We now extend our analysis to broadly assess the role of tidal mixing in the evolution of massive close binary systems. We investigate whether tides enhance or suppress mixing by computing two grids of stellar models, and comparing each time the nitrogen and helium enrichment of the stars in binary systems to that obtained for the single star. We compute twin systems models (see Sect. \ref{binary_interactions}) with initial mass $M= 10,15,20,30,45$ and 60\,M$_\odot$, periods $P=1,1.5,2,3,4,5$ and 7\,days and compare predictions of hydro and magnetic models. For each simulated binary system, we compute a single-star model with identical initial conditions (including the initial velocity).

In the first grid, we assume that tides during the pre-MS phase (not computed) are efficient enough for the stars to be already synchronized at zero-age-main-sequence (ZAMS). This assumption is used in most close binary systems simulations \citep[e.g.,][]{dem09,pax15,mar16,fra23}. Depending on the initial mass and period, this leads to different initial values of $\upsilon_{\rm ini}/\upsilon_{\rm crit}$ that we report in Table \ref{tab:initial_velocities}. This first grid consists of a total of $7\text{ (periods/velocities)}\times 6\text{ (masses)} \times 2\text{ (AMT assumptions)} \times 2 \text{ (binary/single)}= 168$ models.
\begin{table}[h]
\centering
\renewcommand{\arraystretch}{1.3}
\setlength{\tabcolsep}{4pt}
\caption{$\upsilon_{\rm ini}/\upsilon_{\rm crit}$ of the synchronized models at ZAMS. Models filling their RL at ZAMS are indicated with a dash.}
\label{tab:initial_velocities}
\begin{tabular}{!{\vrule width 0.8pt}c!{\vrule width 0.8pt}c!{\vrule width 0.8pt}c!{\vrule width 0.8pt}c!{\vrule width 0.8pt}c!{\vrule width 0.8pt}c!{\vrule width 0.8pt}c!{\vrule width 0.8pt}c!{\vrule width 0.8pt}}
\noalign{\hrule height 0.8pt}
\diagbox[width=6em, height=3.2em,
  linewidth=0.7pt]{$M$ [M$_ \odot$]}{$P$ [days]} & 1 & 1.5 & 2 & 3 & 4 & 5 & 7 \\
\noalign{\hrule height 0.8pt}
60 & -    & - & 0.30 & 0.20 & 0.15 & 0.12 & 0.08  \\
\noalign{\hrule height 0.8pt}
45 & -    & 0.36 & 0.26 & 0.17 & 0.13 & 0.10 & 0.07 \\
\noalign{\hrule height 0.8pt}
30 & -    & 0.31 & 0.23 & 0.15 & 0.11 & 0.09 & 0.06 \\
\noalign{\hrule height 0.8pt}
20 & -    & 0.27 & 0.20 & 0.13 & 0.10 & 0.08 & 0.06 \\
\noalign{\hrule height 0.8pt}
15 & 0.37 & 0.24 & 0.18 & 0.12 & 0.09 & 0.07 & 0.05 \\
\noalign{\hrule height 0.8pt}
10 & 0.32 & 0.21 & 0.16 & 0.10 & 0.08 & 0.06 & 0.04 \\
\noalign{\hrule height 0.8pt}
\end{tabular}
\end{table}
In the second grid, we make the opposite assumption that tides are inefficient during the pre-MS phase. In this case the models (whether single or binaries) are initialized with the typical velocities used in the single-star grids, i.e. $\upsilon_{\rm ini}/\upsilon_{\rm crit}=0.4$. We recall that in the grids by \citet{eks12,cho16}, this value was chosen to reproduce the observed average equatorial velocities of Galactic massive stars \citep{duf06,hua06,eks12} (see also Appendix \ref{AppA}). The second grid consists of a total of $(7 + 1)\text{ (periods\,+\,one single)}\times 6\text{ (masses)} \times 2\text{ (AMT assumptions)}= 96$ models.

\begin{figure}[h]
\centering
\centerline{\includegraphics[trim=0.3cm 0.3cm .3cm 0.3cm, clip=true, width=1\columnwidth,angle=0]{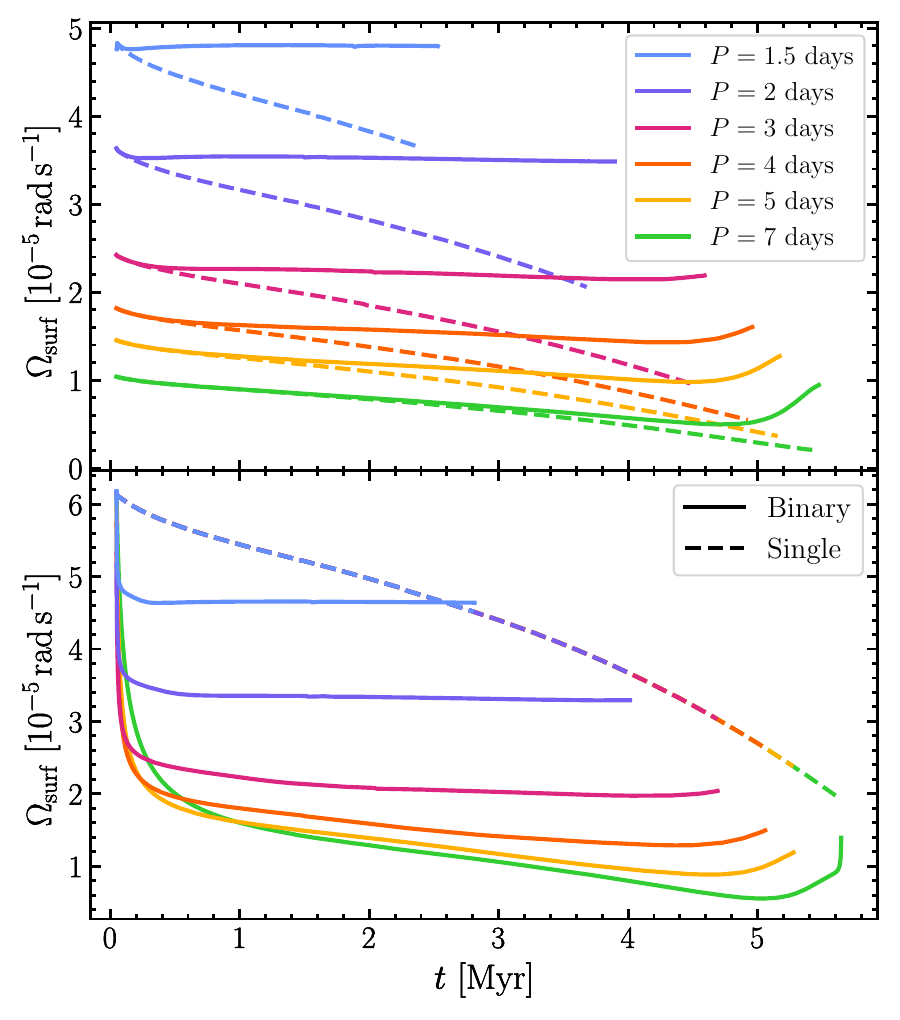}}
\caption{Surface angular velocity evolution of the 30\,M$_\odot$ magnetic models. Binary models are represented in solid lines, single-star models in dashed lines. \textit{Upper panel:} models initialized at synchronization (first grid). \textit{Lower panel:} models initialized at $\upsilon_{\rm ini}/\upsilon_{\rm crit}=0.4$ (second grid).}
\label{omega_evol}
\end{figure}

Fig. \ref{omega_evol} shows the time evolution of the surface angular velocity of the $30$\,M$_\odot$ magnetic models\footnote{The hydro models show a similar surface angular velocity evolution, the main differences are observed in the $\Omega$-profile, since the AMT is less efficient.}. 
With this choice of initial conditions, the binary models of the first grid are accelerated as compared to the single stars (more precisely, the stars are maintained at synchronization, which means that they receive AM from the orbit). Single-star models experience a decrease in $\Omega_{\rm surf}$ with time due to the stellar winds and the MS expansion of the stars. In contrast, when the stars are initialized at $\upsilon_{\rm ini}/\upsilon_{\rm crit}=0.4$ (second grid), binary models are braked until they reach synchronization. Tidal interactions transfer AM from the star to the orbit and binary models generally rotate slower than single-star models. Indeed, with our choice of initial periods and masses, the synchronization velocities at ZAMS are always smaller than $\upsilon_{\rm ini}/\upsilon_{\rm crit}=0.4$ (see Table \ref{tab:initial_velocities}).

In order to quantify whether tides enhance or suppress mixing compared to what is found for the single star, we introduce the quantities $\Delta\widetilde{\text{N}}_{\rm surf}$ and $\Delta \widetilde{Y}_{\rm surf}$, defined as:
\begin{equation}
\begin{split}
    \Delta \widetilde{\text{N}}_{\rm surf}&=\frac{[\text{N}/\text{H}]_{\rm binary}-[\text{N}/\text{H}]_{\rm single}}{X_{0}-X_{\rm c,final}};\\
    \Delta \widetilde{Y}_{\rm surf}&=\frac{Y_{\rm surf,binary}-Y_{\rm surf, single}}{X_{0}-X_{\rm c,final}}.
\end{split}
\end{equation}
The signs of $\Delta\widetilde{\text{N}}_{\rm surf}$ and $\Delta \widetilde{Y}_{\rm surf}$ indicate whether the star experiences more chemical mixing when found in a binary system (positive) or single (negative). The stopping condition is either the reaching of the TAMS (hence $X_{\rm c,final}=0$), or the onset of mass transfer (hence $X_{\rm c,final}=X_{\rm c,RLOF}$). For the binaries, we use the nitrogen and helium enrichment of the last computed model. For the single-star models, the values are taken at the same $X_{\rm{c,final}}$ as the corresponding binary model. The term in the denominator computes how much H has been burnt before the model is stopped, with $X_{0}$ the initial hydrogen mass fraction. This normalization aims at better quantifying the strength of mixing. As the different systems are initialized with different orbital periods, they do not fill their RL at the same moment of their evolution. Longer period systems are wider and thus evolve longer before overfilling the RL. One may conclude that mixing is more efficient in these longer period systems simply because they had more time to evolve, even though at a fixed evolutionary stage (characterized by the central hydrogen mass fraction $X_{\rm c}$), their enrichment are weaker than those of shorter period systems.
\subsubsection{First grid: models initialized at synchronization}\label{grid_init_sync}
\begin{figure*}[h]
\hspace{.75cm}
\begin{subfigure}[b]{0.5\textwidth}
\centering
\centerline{\includegraphics[trim=1.6cm 2.7cm 0.4cm 1.7cm, clip=true, width=0.8\columnwidth,angle=0]{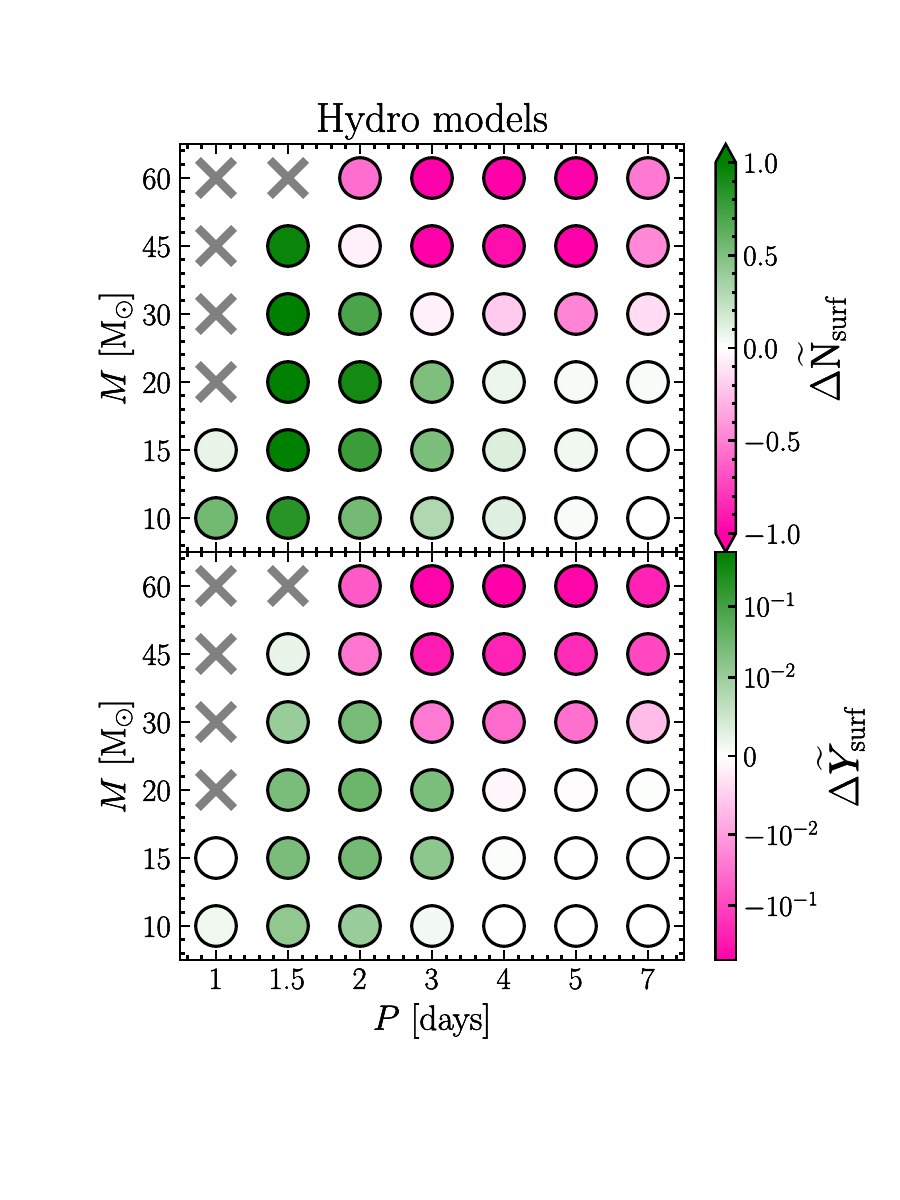}}
\end{subfigure}
\hspace{-1.5cm}
\begin{subfigure}[b]{0.5\textwidth}
\centering
\centerline{\includegraphics[trim=1.6cm 2.7cm 0.4cm 1.7cm, clip=true, width=0.8\columnwidth,angle=0]{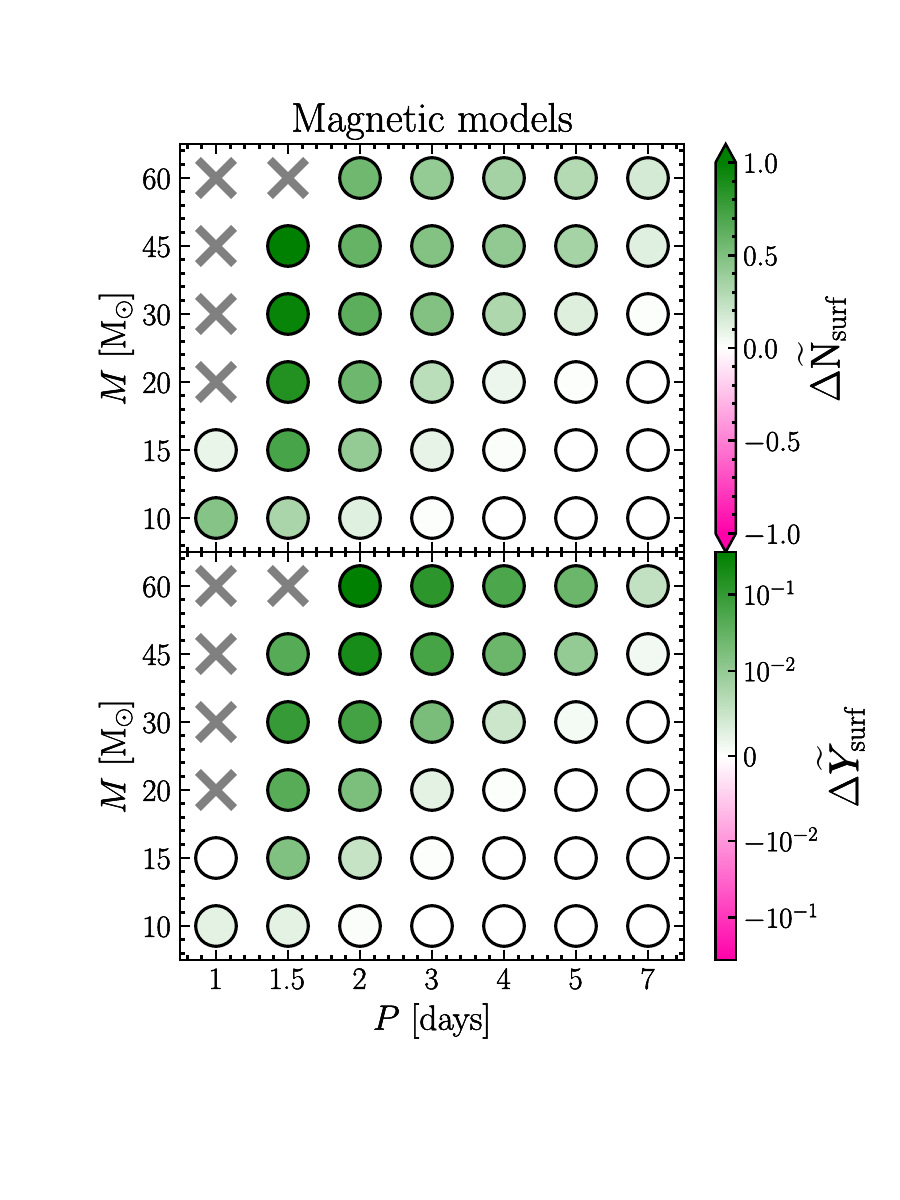}}
\end{subfigure}
\caption{Period-mass diagram comparing the enrichment of binary models initialized at synchronization to those of single-star models with identical initial conditions. The green (magenta) color is used for models with $\Delta>0$ ($\Delta<0$). A symmetric colorbar is used to illustrate the magnitude of the difference in enrichment. Models with $\Delta\sim 0$ appear in white. \textit{Left panels:} hydro models. \textit{Right panels:} magnetic models. \textit{Upper panels:} nitrogen enrichment. \textit{Lower panels:} helium enrichment. Systems overfilling their RL at ZAMS are represented by grey crosses.}
\label{grid_sync}
\end{figure*}

$\Delta\widetilde{\text{N}}_{\rm surf}$ and $\Delta \widetilde{Y}_{\rm surf}$ of the models initialized at synchronization are shown in Fig. \ref{grid_sync}.
Models with $P=1$\,day, $M= 20-60$\,M$_\odot$ and the 60\,M$_\odot$ model at $P=1.5$\,days already fill their RL at ZAMS (gray crosses). In this configuration, binary models are accelerated compared to single-star models (Fig. \ref{omega_evol}). One may thus expect them to be systematically more enriched than single-star models ($\Delta\widetilde{\text{N}}_{\rm surf}>0$ and $\Delta \widetilde{Y}_{_{\rm surf}}>0$). While this is systematically the case for magnetic models, the picture is more complex for hydro models. As illustrated in Sect. \ref{comp_with_song}, this is attributed to the fact that mixing in magnetic models only depends on the value of $\Omega$, whereas in hydro models it also depends on its gradient. By maintaining the stars at synchronization, tides reduce the steepening of the $\Omega$-gradients, which occurs during the MS of the single stars (see Fig. \ref{rapom} in Appendix \ref{AppE}, upper panel). As can be observed in the left panel of Fig. \ref{grid_sync}, the competition of these two effects of tidal torques (increase in $\Omega$, decrease in $\Omega$-gradients) splits the parameter space of ($P,M$) into two different regions. Binary models with low mass and short period are generally more enriched than their single-star counterpart, whereas at high mass and long period, they are generally less enriched than their single-star counterpart. Finally, all magnetic and hydro models with $M<30$\,M$_\odot$ and $P>3$\,days have $\Delta\widetilde{\text{N}}_{\rm surf}\sim 0$ and $\Delta \widetilde{Y}_{_{\rm surf}}\sim 0$, which is explained by their low initial velocities (Table \ref{tab:initial_velocities}). RM is not very efficient, resulting in only minor differences in enrichment between single and binary models.

\begin{figure*}[h]
\hspace{-8pt}
\begin{subfigure}[b]{0.5\textwidth}
\centering
\centerline{\includegraphics[trim=0.07cm 0.298cm 0.36cm 0.312cm, clip=true, width=0.85\columnwidth,angle=0]{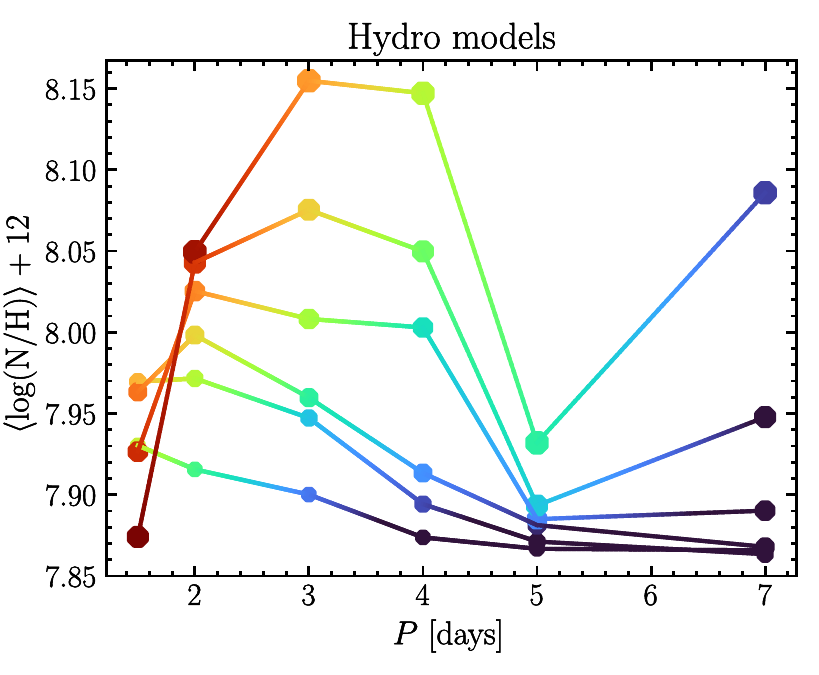}}
\end{subfigure}
\hspace{-13pt}
\begin{subfigure}[b]{0.5\textwidth}
\centering
\centerline{\includegraphics[trim=0cm .3cm 0.7cm 0.3cm, clip=true, width=1\columnwidth,angle=0]{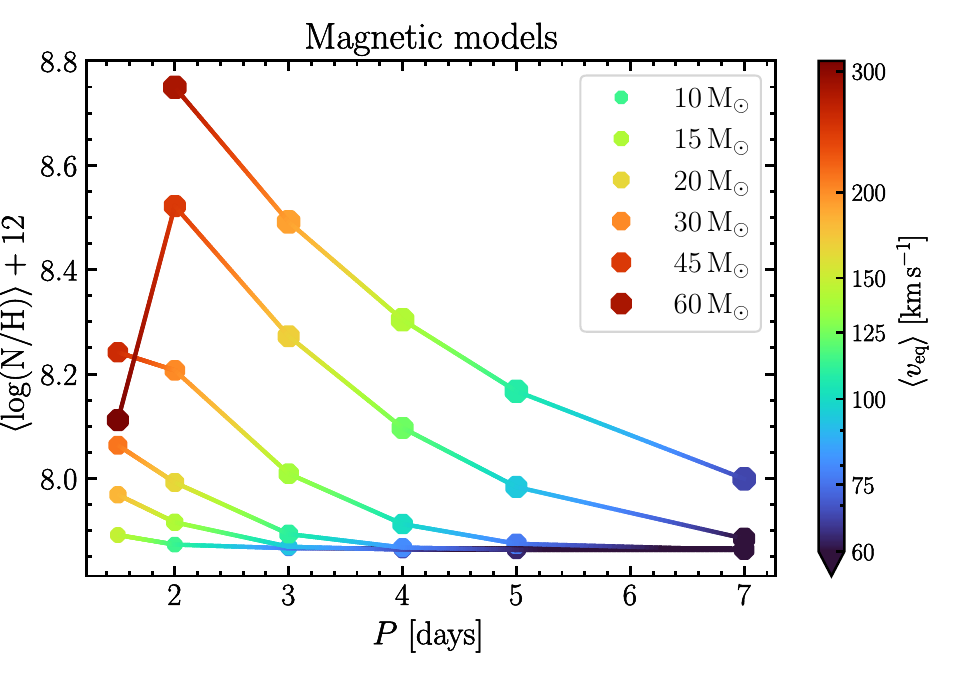}}
\end{subfigure}
\caption{Period-average nitrogen enrichment diagrams across the masses of the grid. Models are color-coded according to their average equatorial velocity. \textit{Left panel:} hydro models. \textit{Right panel:} magnetic models.}
\label{period_enrichement}
\end{figure*}

In order to more clearly illustrate the period dependence of the nitrogen enrichment, we show in Fig. \ref{period_enrichement} the time-averaged nitrogen enrichment of the binary models, defined as
\begin{equation}
    \braket{\log(\text{N}/{\text{H}})}=\frac{1}{t_{\rm final}}\int_{0}^{t_{\rm final}}\log(\text{N}/{\text{H}})\, \text{d}t,
\end{equation}
as a function of period for the considered masses. $t_{\rm final}$ can be $t_{\rm final}=t_{\rm RLOF}$ or $t_{\rm final}=t_{\rm TAMS}$ depending on whether the stars fill their RL during the MS. This average enrichment represents a suitable quantity for comparison with observations. Models are color-coded based on their average equatorial velocity, defined analogously to the average enrichment, highlighting that in synchronized systems, short orbital periods are associated with high rotational velocities.

The magnetic models show a clear period-nitrogen enrichment pattern. The enrichment increases with decreasing period, which can be explained by the fact that at short period models rotate faster. This is consistent with the findings of previous studies \citep[e.g.,][]{yoo06,dem09,man16,mar16,son16}. The low nitrogen enrichment of the 45\,M$_\odot$ model at $P=1.5$\,day is due to the fact that it fills its RL shortly after the beginning of the evolution. We note that RM is expected to continue during a mass transfer phase (or in a contact phase, which is expected to occur at short periods), which would increase the expected average nitrogen enrichment of these close systems. Thus, the predictions of Fig. \ref{period_enrichement} are relevant for pre-interaction systems. The nitrogen enrichment of magnetic models also shows a strong mass dependence, with massive models being more enriched than lower mass ones.

The average nitrogen enrichment of hydro models shows a more complex period and mass dependence. This is mainly due to the reduction of the $\Omega$-gradients by tides, which mitigates or even suppresses mixing. In the mass range $10-30$\,M$_\odot$, the overall period dependence is similar, but less pronounced than in magnetic models. Similarly to what is found for the 45\,M$_\odot$ magnetic model, we notice a drop at $P=1.5$\,day, which is explained by the fact that models at $P=1.5$\,day experience RLOF early, before significant enrichment has time to occur. This effect is more pronounced in hydro models as their enrichment is slower. We also observe that models with $M= 45,60$\,M$_\odot$ are more enriched at $P=7$\,days than at $P=5$\,days. This can explained by two reasons. First, models at $P=7$\, days evolve longer before filling their RL, which increases their time-averaged enrichment. Second, in this high mass regime shear mixing is strong and very sensitive to the degree of differential rotation. At $P=7$\,days, tides are slightly less efficient in maintaining the stars at synchronization (see Fig. \ref{omega_evol}), and therefore $M=45,60$\,M$_\odot$ models have steeper $\Omega$-gradients than at $P=5$\,days, increasing the strength of shear mixing.

Fig. \ref{period_enrichement} illustrates that depending on the chosen AMT assumption, models predict a clear trend in the period-enrichment diagram (magnetic models) of short-period pre-interaction systems, or a more complex period-enrichment dependence (hydro models). Although both types of models predict a mass dependence in the nitrogen enrichment, it is much more pronounced in the magnetic models. Moreover, hydro models generally predict lower nitrogen enrichment than magnetic models, providing further evidence that tides often mitigate mixing in the former. These contrasting results represent key model predictions to be tested against observations. 

Finally, we note that no matter the AMT treatment, the predicted enrichment of binary models with $P>4$\,days are very low. Consistently with the findings of \citet[][their Fig. 3]{dem13}, in this regime tides are efficient and the synchronization velocities relatively low compared to the average velocities used for the calibration of the models (see Table \ref{tab:initial_velocities} and Appendix \ref{AppA}), suppressing RM.
\subsubsection{Second grid: models initialized at $\upsilon_{\rm ini}/\upsilon_{\rm crit}=0.4$}\label{grid_init_nosync}
\begin{figure*}[h]
\hspace{.75cm}
\begin{subfigure}[b]{0.5\textwidth}
\centering
\centerline{\includegraphics[trim=1.6cm 2.7cm 0.4cm 1.7cm, clip=true, width=0.8\columnwidth,angle=0]{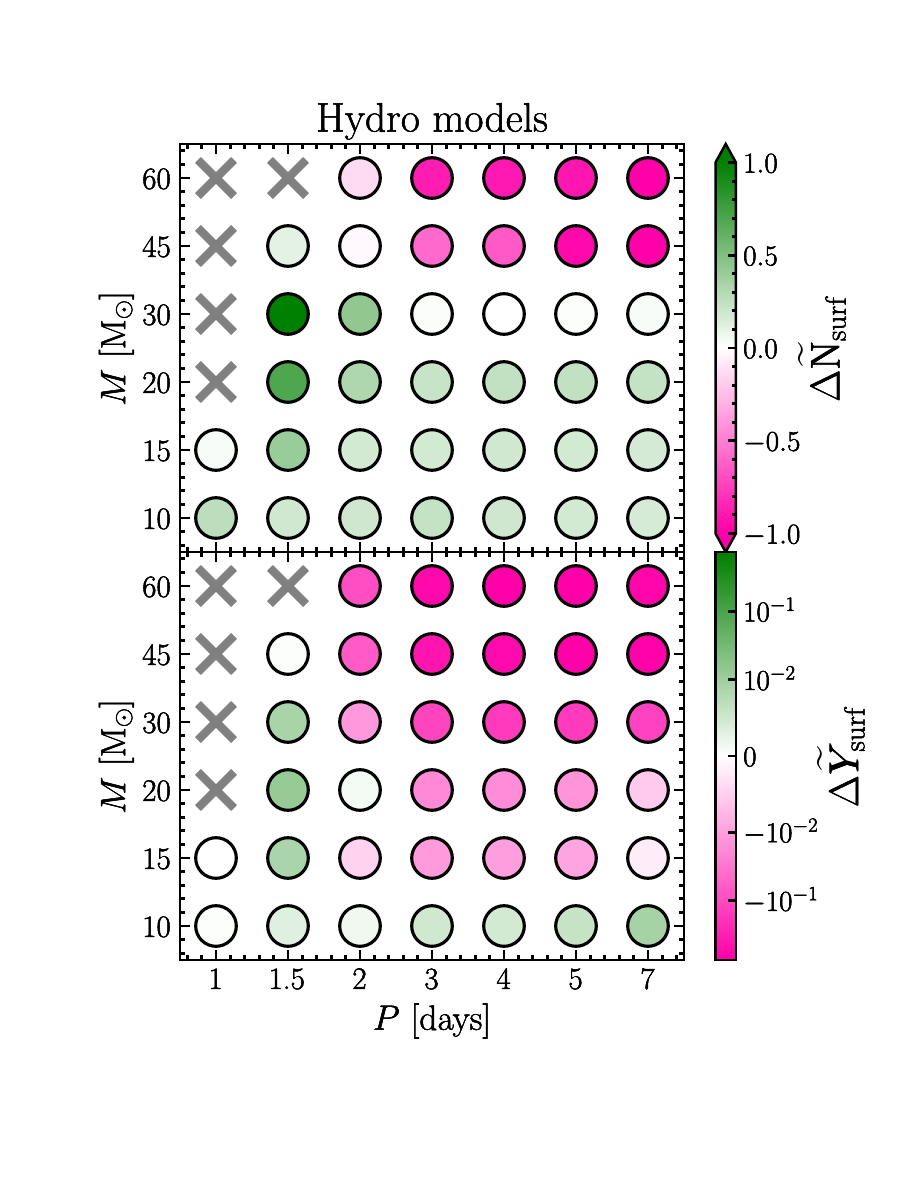}}
\end{subfigure}
\hspace{-1.5cm}
\begin{subfigure}[b]{0.5\textwidth}
\centering
\centerline{\includegraphics[trim=1.6cm 2.7cm 0.4cm 1.7cm, clip=true, width=0.8\columnwidth,angle=0]{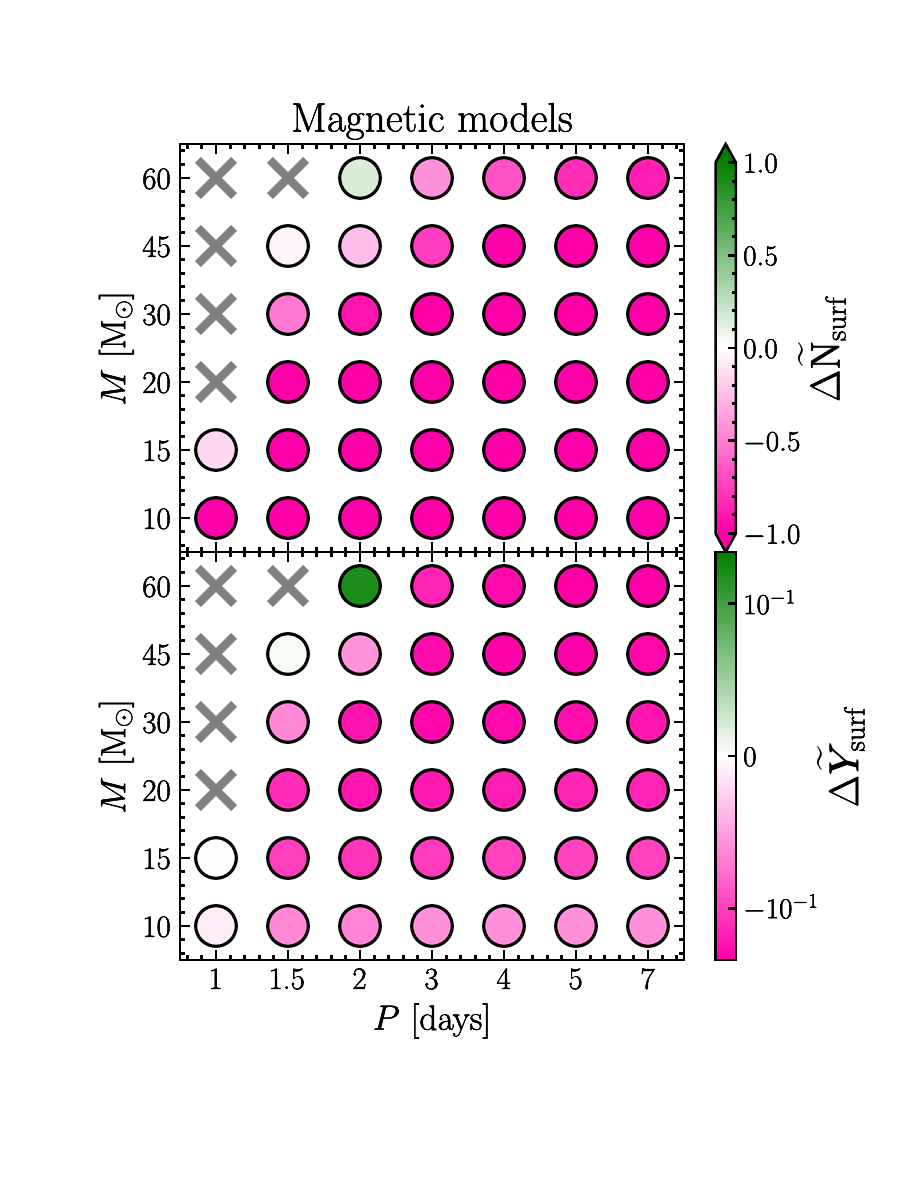}}
\end{subfigure}
\caption{Same as Fig. \ref{grid_sync} for models initialized at $\upsilon_{\rm ini}/\upsilon_{\rm crit}=0.4$.}
\label{grid_nosync}
\end{figure*}

$\Delta\widetilde{\text{N}}_{\rm surf}$ and $\Delta \widetilde{Y}_{\rm surf}$ of the models of the second grid are shown in Fig. \ref{grid_nosync}.
In the chosen mass and period ranges, the synchronization velocities are lower than the value $\upsilon_{\rm ini}/\upsilon_{\rm crit}=0.4$ generally used in the single-star grids (Table \ref{tab:initial_velocities}). As a result, stars in binaries are braked by tides (Fig. \ref{omega_evol}), which reduces the enrichment of magnetic models compared to the single-star counterparts with identical initial conditions.
The only exception is the system with $P=2$\,days, $M=60$\,M$_\odot$. In this case, although the initial velocity is higher than the synchronization velocity of the binary components, the single star slows down due to winds and MS expansion, so that its time-averaged rotational velocity is lower than that of the binary model, which explains the positive values found for $\Delta\widetilde{\text{N}}_{\rm surf}$ and $\Delta \widetilde{Y}_{\rm surf}$. 

The enrichments of binary hydro models show a more complex mass and period dependence. The strong
tidal torques at the beginning of the evolution increase the $\Omega$--gradients, boosting the nitrogen transport, which explains the positive sign of $\Delta\widetilde{\text{N}}_{\rm surf}$ for the $10-30$\,M$_\odot$ models (see also Fig. \ref{rapom} in Appendix \ref{AppE}, lower panel). Consistently with the results of Sect. \ref{spin_down}, the tides are less efficient in enhancing helium, which explains why some models have $\Delta \widetilde{\text{N}}_{\rm surf}>0$ but $\Delta \widetilde{Y}_{\rm surf}<0$. In the mass range $45-60$\,M$_\odot$, single-star models are more enriched than binary models, as in Sect. \ref{song_massive}.

\begin{figure*}[h]
\hspace{-8pt}
\begin{subfigure}[b]{0.5\textwidth}
\centering
\centerline{\includegraphics[trim=0cm 0.29cm 0.cm 0.cm, clip=true, width=0.857\columnwidth,angle=0]{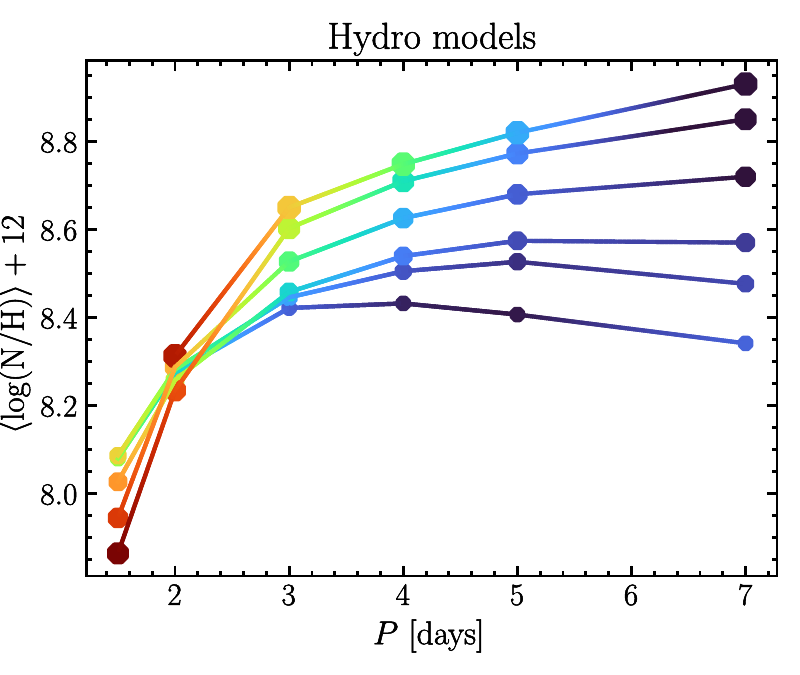}}
\end{subfigure}
\hspace{-18pt}
\begin{subfigure}[b]{0.5\textwidth}
\centering
\centerline{\includegraphics[trim=0cm .29cm 0.7cm 0.31cm, clip=true, width=1\columnwidth,angle=0]{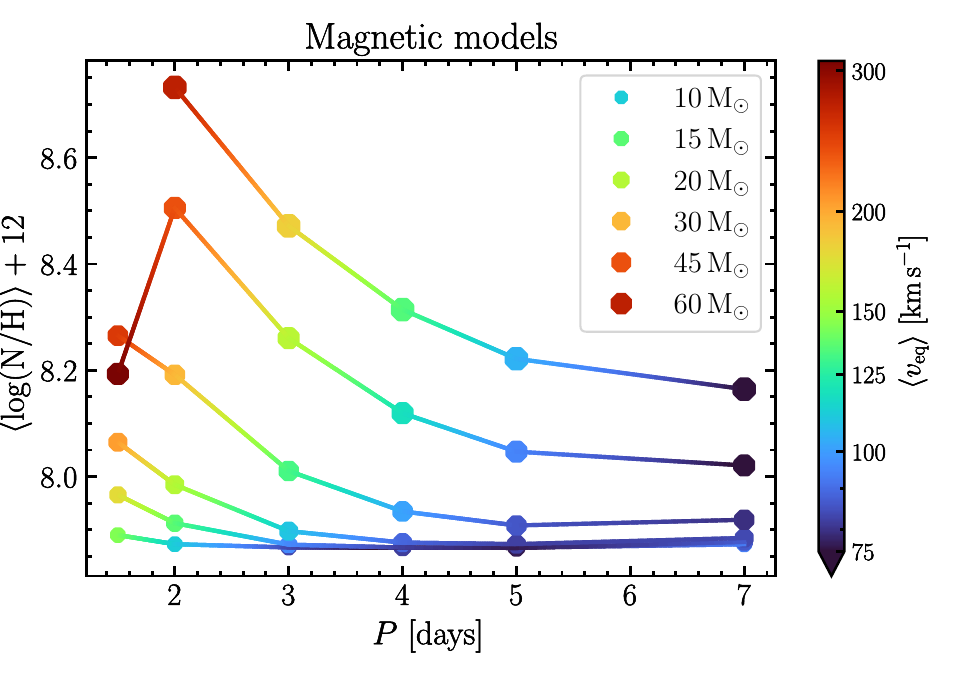}}
\end{subfigure}
\caption{Same as Fig. \ref{period_enrichement} for models initialized at $\upsilon_{\rm ini}/\upsilon_{\rm crit}=0.4$.}
\label{period_enrichement_nosync}
\end{figure*}

The period-average nitrogen enrichment diagrams of the second grid are shown in Fig. \ref{period_enrichement_nosync}. Opposite trends between magnetic and hydro models are observed in this case. In hydro models, the average nitrogen enrichment generally increases with the period, consistently with the results of Sect. \ref{spin_down} and \ref{song_massive}. This can be explained by the dominating effect of shear mixing, which is boosted by the larger degree of differential rotation imposed by the tidal torques at longer periods and the fact that longer period systems evolve longer before interacting. The model predictions significantly differ from those of Sect. \ref{grid_init_sync}, highlighting their sensitivity to the initial conditions.

In contrast, binary magnetic models' predictions are insensitive to the initial velocity: the period-enrichment diagram is almost identical to that of the first grid. This can be explained by the fact in magnetic models, tidal torques quickly synchronize the whole star (see Fig. \ref{profiles} and \ref{omega_profiles}) and mixing only depends on the value of $\Omega$. The general trend in the period-enrichment diagram is the exact opposite than that predicted by the hydro models in this case: the enrichment decreases with the period. Period-enrichment diagrams can be interpreted as velocity-enrichment diagrams (originally introduced by \citealt{hun09,bro11b}), where a long period corresponds to a low velocity, as also illustrated in Figs. \ref{period_enrichement} and \ref{period_enrichement_nosync}. Interestingly, spin-down systems simulated with a hydrodynamic AMT predict higher enrichments for shorter periods, in contrast with standard single-star models, where faster rotators typically exhibit stronger enrichment. We discuss in Sect. \ref{comp_observations} how confronting these predictions to observations of close pre-interacting systems may lend support to or argue against particular types of AMT.
\subsection{Structural reaction to chemical mixing}\label{CHE}
In the most massive stars of the grids presented in Sect. \ref{grid}, RM is strong enough for significantly altering the surface abundance of helium, which induces substantial structural changes. The helium enhancement lowers the opacity of the surface layers, which prevents the stars from expanding, or even makes them contract \citep[e.g.,][]{mey00,dem09,son16}. In extreme cases this effect can lead to unusual evolutionary pathways, such as a mass transfer initiated by the secondary star \citep{dem09} or the so-called chemically homogeneous evolution \citep[e.g.,][]{man16,mar16}, where both components stay within their RL and mass transfer is avoided.

\begin{figure*}[h]
\centering
\centerline{\includegraphics[trim=0.35cm 0.2cm 0.35cm 0.35cm, clip=true, width=1.5\columnwidth,angle=0]{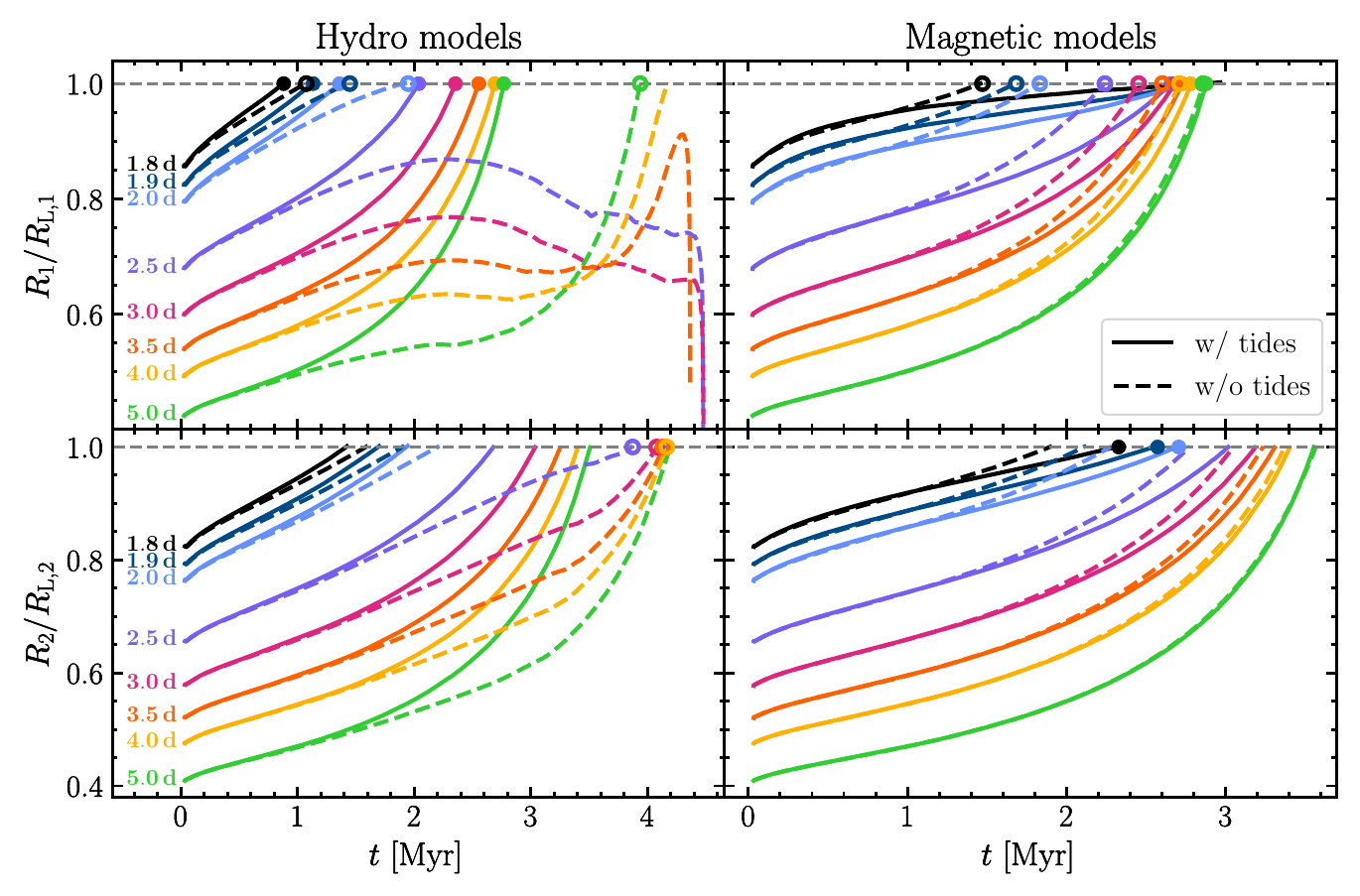}}
\caption{Roche lobe filling factor evolution of 60+45\,M$_\odot$ systems at different orbital periods. Models accounting (not accounting) for tidal interactions are represented in solid (dashed) lines.
\textit{Left panels:} hydro models: \textit{Right panels:} magnetic models. \textit{Upper panels:} primary star evolution. \textit{Lower panels:} secondary star evolution. For each simulated system a dot indicates whether the primary or the secondary fills its RL first (solid dot for the models with tides, empty dot for the models without tides).}
\label{rl_filling_factors}
\end{figure*}

In Fig. \ref{rl_filling_factors}, we show the evolution of the Roche lobe filling factors $R/R_{\rm L}$ of 60 + 45\,M$_\odot$ systems at solar metallicity in the period range $1.8-5$\,days. Systems are initialized at synchronization, therefore tides transfer AM from the orbit to the spin, as in Sect. \ref{grid_init_sync}. In this case, we successively simulated the evolution of the primary and the secondary, and stopped the simulations at RLOF, as was done in \citet{dem09} for SMC models. We compare the evolution predicted by hydro and magnetic models. We also include model variations where the stars are initiated with identical initial conditions but that do no account for the tidal torques. This allows us to assess whether tides increase or suppress mixing as compared to the single-star evolution. For these model variations, we also stop the simulations at RLOF.

When the tidal torques are neglected and the models behave like single stars, the primaries of the hydro models with $P=2.5-4$\,days experience strong mixing, consistently with the results of Sect. \ref{grid_init_sync}, which makes them contract after reaching a maximum radius smaller than the RL radius. As a result, the secondaries fill their RL first. Magnetic models neglecting tides experience little mixing and in this case the primaries systematically fill their RL first. In contrast, when the tidal torques are accounted for, in hydro models the primaries systematically fill their RL first. In magnetic models, the primaries of the $P=1.8-2$\,days systems experience strong mixing, which slows their expansion and makes the secondaries fill their RL first. This highlights opposite models' chemical and structural reactions to tidal torques depending on the AMT assumptions.
\section{Discussion}\label{discussion}
\subsection{Comparison with previous studies}
The primary conclusion of the studies by \citet{son13,son16} is that tidal interactions in binary models consistently enhance mixing as compared to single-star models with identical initial conditions. In this paper, we expand on this finding by testing its robustness across various system configurations and AMT treatments. By reproducing the systems presented in \citet{son13} with different AMT assumptions, we show that in spin-down configurations, mixing is generally suppressed in binary magnetic models. The picture is different in hydro models where the spin-down boosts shear mixing. This may lead the stars in binary systems to experience more mixing than the single-star counterpart, although they synchronize at low angular velocities (Sect. \ref{spin_down}). We demonstrate that this is not a general result, as there also exists spin-down configurations where binary hydro models predict less enrichment than what is found for the single star (Sect. \ref{song_massive} and \ref{grid_init_nosync}). Additionally, in Sect. \ref{grid_init_sync} we highlight that when stars are initiated at synchronization, binary hydro models may experience less mixing than the single star with identical initial velocity, although tides in this case increase their AM content. Finally, the conclusions may change depending whether helium or nitrogen is considered as tracer of RM. Although both are enhanced by the CNO process, the mixing of the processed element not only depends on the diffusion coefficients --which are the same for all species-- but also on the abundance profiles themselves. The fact that nitrogen is initially less abundant than helium implies that it is more efficiently mixed at early stages of evolution, as a strong chemical gradient is quickly established.

In the study by \citet{dem09}, models were initiated at synchronization, assuming that tides are efficient during the pre-MS phase. The AMT treatment is similar to the magnetic models of the present study and imposes a strong core-envelope coupling. In this case, shear mixing is inefficient, and the authors found a clear anti-correlation between period and nitrogen enrichment, which we also obtain in magnetic models. In contrast, hydro models do not show such a clear anti-correlation (Fig. \ref{period_enrichement}). Moreover the trend depends on the assumed initial velocity, unlike in magnetic models (Fig. \ref{period_enrichement_nosync}). In Sect. 5.2 of their paper, \citet{dem09} study the evolution of 50+25\,M$_\odot$ short-period systems at SMC metallicity. At very short periods ($P<1.8$\,days), the primary experiences enough mixing for undergoing substantial structural changes, which makes it stay compact long enough for the secondary to fill its RL first. We find similar results with magnetic models at solar metallicity for 60+45\,M$_\odot$ systems. However, in this mass range hydro models maintained at synchronization experience little mixing and as a result the primary fills its RL first across the whole period range, highlighting the importance of the AMT efficiency in determining the evolutionary pathways of close massive binary systems.
\subsection{Comparison with observations}\label{comp_observations}
The identification of the main process(es) responsible for chemical mixing is crucial and still heavily debated. RM is one of the most promising explanations, but it faces several challenges. \citet{hun09,bro11b} proposed a direct test to the predictions of RM by inspecting correlations between the observed nitrogen enrichment of massive stars and their surface projected rotational velocity. Although the majority of stars in their sample have velocities and nitrogen enrichment compatible with models accounting for RM, they identified two categories of stars challenging the predictions: slow rotators with strong enrichments at relatively high surface gravity (typically reached during the MS phase) and fast rotators with low enrichments. \citet{dem09} expanded on their idea by proposing an equivalent test in massive close binaries, where tidal synchronization is expected to boost RM, so that any system with $P\le3$\,days is expected to show significant nitrogen enrichment. Since the synchronization velocity inversely scales with the period, one expects to find an anti-correlation between period and nitrogen enrichment. Several spectroscopic studies of close binary systems have challenged these predictions. The majority of short-period systems with orbital periods smaller than 10\,days in the studies by \citet{mar17,pav18,pav23,abd19,abd21} do not show any evidence of strong enrichment. None of these studies report any clear anti-correlation between period and nitrogen enrichment; however, this may be attributable to insufficient sample sizes.

\citet{mah20} determined the surface abundances of the 31 double-lined spectroscopic binaries in the Tarantula Massive Binary Monitoring sample, in which at least one of the components is classified as an O-type star. Their subsample 3 is composed of 18 stars in short-period ($P<10$\,days) circular systems. The authors found a trend between the nitrogen enrichments and the projected rotational velocities, the three stars with the highest nitrogen enrichment rotating faster than 150\,km\,s$^{-1}$ at relatively short periods ($P<2$\,days). On the other hand, the remaining stars in their sample do not exhibit any evidence of nitrogen enrichment, despite some having rotational velocities up to 200\,km\,s$^{-1}$, thereby diluting the overall trend.

The tension between observations and predictions of stellar models led \citet{pav18} to wonder about "the role of binarity in terms of tides on damping internal mixing in stars residing in binary systems". The magnetic models presented in this study incorporate physics similar to that of the models of \citet{dem09}, and also predict a clear period-nitrogen enrichment anti-correlation. In this sense, they are also challenged by the observations. In contrast, hydro models offer a larger variety of predictions, as the period-enrichment trends also depend on the choice of initial velocities. When tides are assumed to be efficient during the pre-MS, the models do not predict any clear period-enrichment anti-correlation. Moreover, the time-averaged enrichments are lower than in magnetic models\footnote{Although both types of models were calibrated with the same observables, see Appendix \ref{AppA}.}, especially at high masses and short periods. The lower enrichment in binary hydro models is due to the competing effects of tidal torques on shear mixing: in this configuration they increase the AM content of the stars, but also decrease their $\Omega$-gradients. When tides are assumed to be inefficient during the pre-MS, hydro models even predict a correlation between period and enrichment-- exactly opposite to the predictions of magnetic models. This prediction would be compatible with observed systems at intermediate periods ($5$\,days $\lesssim P\lesssim 20$\,days)\footnote{The efficiency of tides is expected to rapidly decrease above $\sim10-20$\,days.} or apparently single stars with an undetected companion, rotating at low velocities due to the spin-down by tides and showing large nitrogen enrichments.

We stress that the efficiency of tides during the pre-MS of massive binaries is extremely uncertain. Observational studies that robustly assess tidal synchronization during the pre-MS are lacking so far \citep{len24}. On the theoretical side, \Genec pre-MS single-star models with $M\ge 3$\,M$_\odot$ by \citet{hae19} all undergo a fully-radiative phase before the onset of H-burning. The dynamical tides considered in this study, as formulated by \citet{zah77}, are expected to be completely inefficient in fully radiative stars. Consequently, one may argue that the ZAMS rotation rates of massive close binaries are more likely set by their pre-MS contraction phase and other processes at play, i.e., accretion disks, magnetic fields and jets \citep[e.g.,][]{sei11,pud19,com22,oli23} than by tidal interactions, unless we are missing an additional, efficient dissipation mechanism.
\subsection{Main caveats}\label{caveats}
The main limitations of the models presented in this work are the following:
\begin{enumerate}
    \item The value of $\alpha_{\rm ov}$ used in \citet{eks12} grid and the present works may be underestimated for stars above $9$\,M$_\odot$ \citep[e.g.,][]{cas14,tka20,mar21,bar23}. For the sake of the comparison with the results of \citet{son13} we considered necessary to adopt consistent stellar physics inputs as we mainly focused on the interplay between tides and RM in this study. We note that the results of the magnetic models are overall in agreement with those of \citet{dem09}, who used $\alpha_{\rm ov}=0.355$.
    \item The models do not directly account for mixing nor AMT by IGWs. \citet{zah77} formalism of dynamical tides is based on the assumption of IGWs traveling from the convective-core boundary to a region close to the surface, where they dissipate energy. Theoretically IGWs can also transport AM \citep[see e.g.,][]{tal03,rog13,rog25} and chemicals \citep[e.g.,][]{mon94,mon96,her23,bri24,bri25,var24,mom25}, which was neglected in this work. However, a proper treatment of tidally excited IGWs is challenging to implement, and to date, thorough investigations of the interplay between RM and AMT accounting for the transport by tidally excited IGWs are lacking. \citet{tal98} studied the joint action of meridional circulation, shear, and tidally excited IGWs in the AMT transport of $9$\,M$_\odot$ models, but did not discuss the resulting impact on RM. Recent studies by \cite{var24,mom25,bri25} compare the efficiency of mixing by stochastically excited IGWs and rotation. They conclude that mixing by IGWs offers a promising explanation to the "outliers" of the Hunter diagram, as this process is not incompatible with high enrichment at low rotations. \citet{var24} even find the mixing efficiency to decrease with rotation, providing an explanation for fast rotators with low enrichment. Interestingly, the models presented in this study display a large variety of trends, offering alternative (and not incompatible) explanations, while relying on the physics of RM. It should be noted that in binaries, the competition between stochastically and tidally excited IGWs further complicates the picture.
    \item Resonant locking is not considered in this study, which may alter the synchronization velocities of the models \citep[e.g.,][]{wit02,bur12,ma21,fel25}, and in turn the efficiency of RM.
    \item The models are stopped at RLOF, as we can not follow mass transfer and contact phases in \Genec at this stage. We therefore caution that the period-enrichment diagrams (Figs. \ref{period_enrichement} and \ref{period_enrichement_nosync}) are relevant for pre-interacting systems.
    \item We restricted our analysis to solar-metallicity models. Extending the study to lower metallicities would be of particular interest, as they influence both mass-loss rates and RM. We plan to pursue this in future work.
\end{enumerate}
\section{Conclusion}\label{conclusion}
In this study we computed several grids of single stars and close, pre-interacting binary systems with \Genec in the mass and period range $10-60$\,M$_\odot$, $1-7$\,days. We took benefit of the versatility in rotation treatment of the \Genec code to study the interplay between tidal interaction and rotational mixing with various angular momentum transport assumptions. We investigated whether tidal interactions enhance or suppress mixing by computing single-star model variations with identical initial conditions. We also proposed a refined treatment of tidal interactions, inspired from \citet{fra23}.

We find that when a strong AMT is accounted for (magnetic models with calibrated Tayler-Spruit dynamo), stars rotate nearly as solid-bodies, making shear mixing inefficient. In this case, the efficiency of rotational mixing is only determined by the angular velocity of the stars. As a result, the impact of tidal interactions on chemical mixing comes down to the question of whether tides spin up or spin down the stars. The corollary of this result is that mixing in binary magnetic models is rather insensitive to the assumed initial velocity and is only determined by the system configuration, as long as the synchronization timescale is short with respect to the nuclear timescale.

The picture is more complex in hydro models, which are subject to shear mixing due to their moderate core-envelope coupling. In this case, the efficiency of rotational mixing also depends on the angular velocity gradients. This intrinsic difference of the models has major consequences on their reaction to tidal torques. Spin-down models may experience more mixing than single stars with identical initial conditions, although tides make them rotate at lower angular velocities. Conversely, binary models maintained at synchronization may experience less mixing than their single-star counterparts. Contrary to magnetic models, in hydro models the efficiency of rotational mixing not only depends on the system configuration, but also on the assumed rotational velocity at ZAMS.

These disparities have multiple consequences. First, the predicted evolutionary pathways of massive close binary systems may differ. The results of Sect. \ref{CHE} indicate that the chemically homogeneous evolution \citep[][]{dem09,man16,mar16} is likely more difficult to achieve with hydro models. Thus, we expect the predicted merger rates to sensibly depend on the considered AMT, even though the RM efficiency of both types of models are calibrated with the same observables (see Appendix \ref{AppA}). Second, the predicted period-enrichment diagrams --suitable for direct comparison with observations-- significantly differ. While magnetic models predict a clear anti-correlation between period and nitrogen enrichment, the picture is less clear in hydro models, and the predictions strongly depend on the ZAMS velocity. Notably, when tides are assumed to be inefficient during the pre-MS, the latter even predict a correlation between period and enrichment.

Recent observations challenge the predictions of models accounting for rotational mixing, as numerous massive close binaries rotating at high velocities do not show any evidence of nitrogen enrichment \citep[e.g.,][]{mar17,pav18,pav23,abd19,abd21}. Conversely, some longer period systems in the sample by \citet{mah20} show larger nitrogen enrichment than systems in tighter orbits. No clear anti-correlation between period and enrichment is reported in these studies. Consistently with the findings of \citet{dem09}, our magnetic models show a clear period-nitrogen anti-correlation, and are thus challenged by these observations. Without claiming that these observations strongly favor hydro models, we note that they better stand the comparison, owing to the greater versatility of their predictions --stemming from their higher sensitivity to initial conditions. The need for statistically significant observational constraints on RM remains critical-- a gap that current and upcoming large-scale surveys, such as IACOB \citep[][]{sim14,deb24,mar25} and BLOeM \citep{she24}, will help to narrow.
\section{Data availability}
The tracks are available on \href{https://zenodo.org/records/18302392}{Zenodo}, \href{https://doi.org/10.26037/yareta:gejbckay45bjhnrzklcev6tb6u}{Yareta}, and on the \href{https://www.unige.ch/sciences/astro/evolution/en/database}{Geneva stellar group database}.
\begin{acknowledgements}
LS, SR and SE have received support from the SNF project No 212143. MM and PE have received support from the SNF project No 219745.
\end{acknowledgements}

\bibliographystyle{aa}
\bibliography{myrefs}\begin{appendix}
\section{Calibration of the magnetic models}\label{AppA}
In this section, we present the calibration of RM performed for the magnetic models. For this calibration to be consistent with that of the hydro models, the rotational velocities must also be calibrated to reproduce the same observations.

\begin{figure}[h]
\centering
\centerline{\includegraphics[trim=1.6cm 1.3cm 1.6cm 1.5cm, clip=true, width=1.02\columnwidth,angle=0]{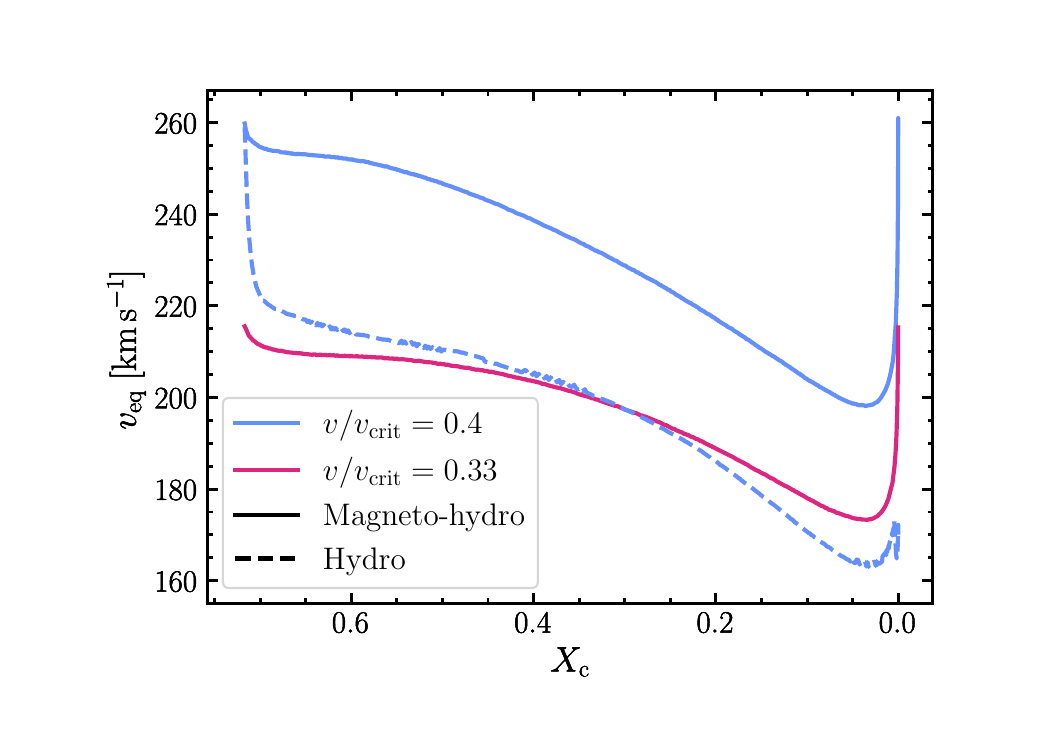}}
\caption{MS evolution of 15\,M$_\odot$ stars with different initial velocities and AMT treatments.}
\label{vel_evol_cal}
\end{figure}

Fig. \ref{vel_evol_cal} shows the MS evolution of the equatorial velocity of 15\,M$_\odot$ single-star models. 
The evolution of two magnetic models and one hydro model are shown. One of the magnetic model and the hydro model are initiated at initial velocity $\upsilon_{\rm ini}/\upsilon_{\rm crit}=0.4$. The other magnetic model is initiated at $\upsilon_{\rm ini}/\upsilon_{\rm crit}=0.33$. Soon after ZAMS, the velocity of the hydro model drops from $\sim 260$\,km\,s$^{-1}$ to $\sim 220$\,km\,s$^{-1}$. This is a consequence of the advective AMT by meridional circulation, which transports AM from the surface to the core (models are initiated with solid body rotation at ZAMS). Magnetic models with a purely-diffusive AMT do not show the same behavior and hence the average equatorial velocity along the MS evolution is higher than that of the hydro model with the same initial value of $\upsilon_{\rm ini}/\upsilon_{\rm crit}$. The value $\upsilon_{\rm ini}/\upsilon_{\rm crit}=0.4$ was chosen so that hydro models reproduce the observed average equatorial velocities of Galactic massive stars \citep{duf06,hua06,eks12}. In order to ensure that the magnetic models also reproduce this constraint, we consider that the initial value of $\upsilon_{\rm ini}/\upsilon_{\rm crit}$ required for performing the calibration should be slightly reduced. We found that $\upsilon_{\rm ini}/\upsilon_{\rm crit}=0.33$ is a satisfactory choice as the average MS equatorial velocity of the 15\,M$_\odot$ magnetic model in this case matches that of the hydro model with $\upsilon_{\rm ini}/\upsilon_{\rm crit}=0.4$.

\begin{figure}[h]
\centering
\centerline{\includegraphics[trim=2.3cm 1.1cm 1.8cm 1.4cm, clip=true, width=1.02\columnwidth,angle=0]{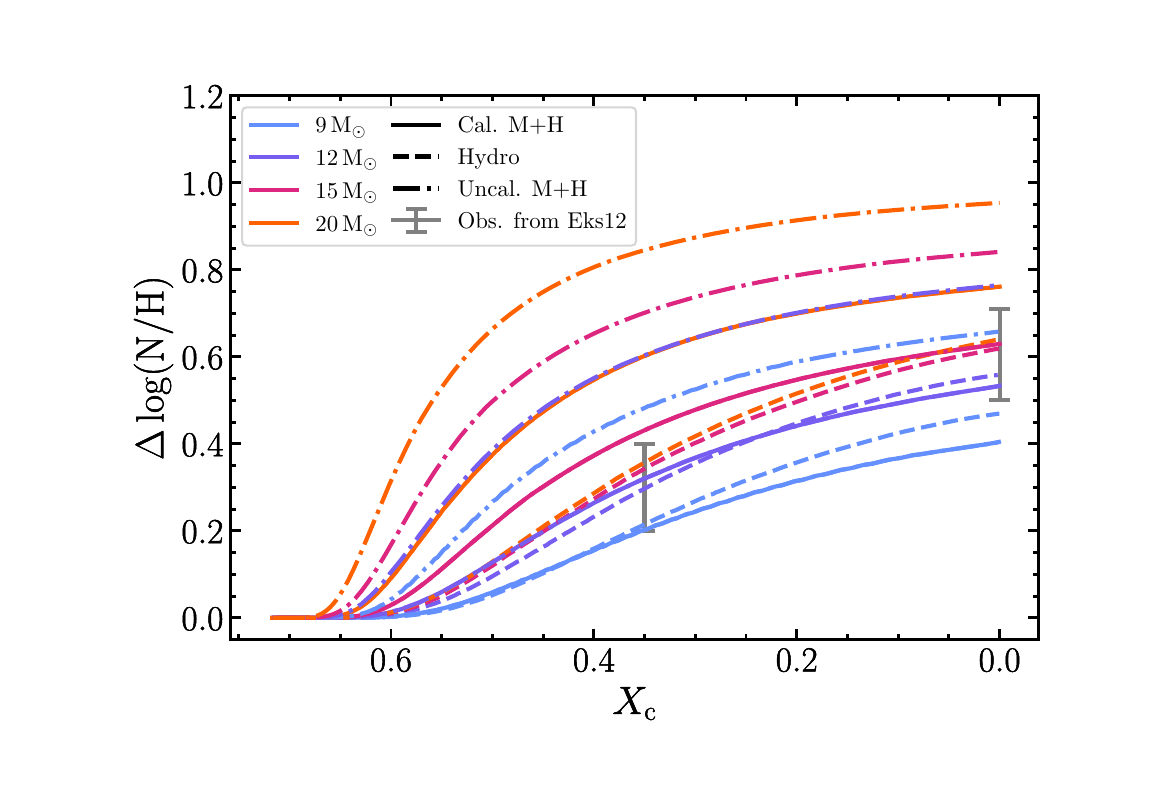}}
\caption{MS nitrogen enrichment of stars in the mass range $9-20$\,M$_\odot$ with different AMT assumptions and RM efficiencies. Hydro models are shown in dashed lines, calibrated magnetic models in solid lines, uncalibrated magnetic models in dot-dashed lines. The gray boxes are compiled observations from \citet{eks12}.}
\label{evol_cal}
\end{figure}

Then, the RM efficiency of the magnetic models is calibrated by tuning the value of $c_{\rm h}$, which is a free parameter in the expression of $D_{\rm h}$. We use the nitrogen enrichment as tracer of RM. We follow \citep{eks12} and use observed nitrogen abundances of Galactic MS B-type stars with initial masses up to 20\,M$_\odot$ from \citep{gie92,kil92,mor08,hun09}. Combining these observations, \citet{eks12} provided two boxes of nitrogen enrichment $\Delta\log($N$/$H$)$ (at mid and end-MS) that models should reproduce. Fig. \ref{evol_cal} shows the MS evolution of $\Delta\log($N$/$H$)$ for \Genec models with initial masses $M=9,12,15$ and 20\,M$_\odot$. Three sets of models are shown: hydro models identical to those in \citet{eks12}, magnetic models without calibration, i.e. with $c_{\rm h}=1$ and calibrated magnetic models with $c_{\rm h}=0.70$.
Uncalibrated magnetic models fail at reproducing the observations, as the mid-MS enrichment of the $9\,$M$_\odot$ model and all the enrichments of the 12-20$\,$M$_\odot$ models are outside the boxes. The calibrated magnetic models show a better agreement with the observations, yet less satisfactory than the hydro models. The first reason is that the enrichment of the magnetic models occurs mainly at the beginning of the MS evolution, whereas hydro models show a more prolonged enrichment. This is explained by the fact that in the latter models mixing depends on both $\Omega$ and its gradient. The MS expansion and AM losses reduce the surface angular velocity of the models, but also steepens the $\Omega$--gradients due to the moderate AMT, which sustains shear mixing. In magnetic models, shear mixing is inefficient and the enrichment is suppressed when the angular velocity decreases. As a result, it is harder to match both the mid- and the end-MS enrichments with magnetic models. The value of $c_{\rm h}$ was chosen so that a maximum number of models reproduce the observations across the mass range. With this calibration, the $9\,$M$_\odot$ model just enters the box at mid-MS. At end-MS, it is however lower than the box. In contrast, the 20\,M$_\odot$ model is too enriched and lies outside the box both at mid- and end-MS. The 15\,M$_\odot$ model enters the box at end-MS but not at mid-MS. This highlights that RM in magnetic models exhibits a stronger mass dependence than in hydro models, which is the second reason why magnetic models overall reproduce the observed nitrogen abundances less satisfactorily.
\section{Evolution towards synchronization with and without equilibrium tides}\label{AppB}
\begin{figure}[h]
\centering
\centerline{\includegraphics[trim=1.9cm 3.cm 1.8cm 2.4cm, clip=true, width=1.02\columnwidth,angle=0]{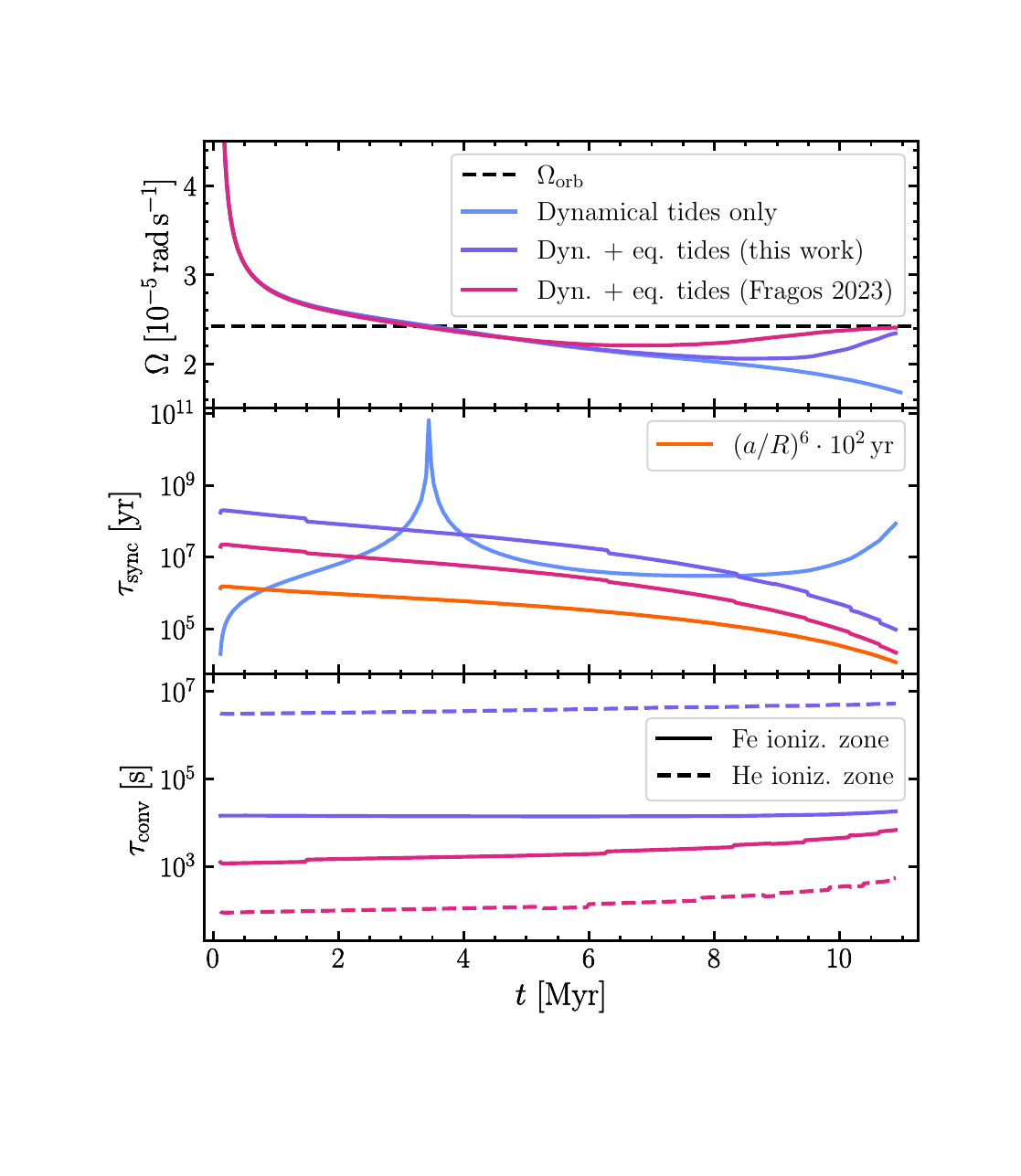}}
\caption{Evolution towards synchronization of the same system as in \cite{sci24} with $P=3$\,days including model variations accounting for the equilibrium tides.
\textit{Upper panel:} angular velocity. \textit{Middle panel:} synchronization timescales (the different lines correspond to the dynamical tides and equilibrium tides synchronization timescales, not the "global" synchronization timescales of each computed model). \textit{Lower panel:} convective turnover timescales.}
\label{tides_method}
\end{figure}

We show in Fig. \ref{tides_method} the evolution towards synchronization of the same system as in \citet{sci24} with an initial period of $P=3$\,days. The evolution obtained with only dynamical tides is compared to that including equilibrium tides, treated either with the prescription adopted in this work (Eq. \eqref{prescription}) and that of \citep[][i.e, Eq. \eqref{posydon}]{fra23}. This illustrative example highlights the importance of equilibrium tides in the tidal evolution and allows us to assess the relative strength of the present prescription against that of \citet{fra23}\footnote{For any choice of equilibrium tide prescription, the dynamical tides were modeled according to Eq. \eqref{correct} (which is not the case of the \citealt{fra23} models who use the \citealt{hur02} adaptation of the \citealt{zah77} prescription).}. Models are stopped at RLOF.

At the beginning of the simulation, the dynamical tides dominate the evolution as the three lines overlap until $t\sim 6$\,Myrs. At the end of the evolution, the equilibrium tides dominate, as shown by the evolution of the synchronization timescales, defined as $\tau_{\rm sync}=\left|\frac{\Omega_{\rm spin} - \Omega_{\rm orb}}{\dot \Omega_{\rm spin}}\right|$. As a result, the models accounting for equilibrium tides are closer to synchronization at RLOF. The relative departure from synchronization $\left|\frac{\Omega_{\rm spin} - \Omega_{\rm orb}}{\Omega_{\rm orb}}\right|$ is 0.66\% for the model computed with the \citet{fra23} prescription, 3\% for the model computed with the updated prescription of this work, and 31\% for the model accounting only for the dynamical tides. The large departure from synchronization at RLOF of the model without equilibrium tides can be imparted to the decrease of $E_2$ along the MS and the power 8/3 in $\Omega_{\rm orb}-\Omega_{\rm spin}$ of the dynamical tides torque (Eq. \eqref{correct}), which makes it very inefficient close to synchronization. Equivalently, the dynamical tides synchronization timescale diverges at synchronization, which can be seen at $t\sim 3.4$\,Myrs, when the star "crosses" the synchronization. Unlike the equilibrium tides, the dynamical tides synchronization timescale depends on $\Omega_{\rm orb}-\Omega_{\rm spin}$ (Eq. (A.1) in \citealt{sci24}), which explains the significantly different evolution the timescales follow. The evolution of the equilibrium tides synchronization timescales follow the same trend as $(a/R)^6$, which is shown as a reference.

Under the formulation by \citet{fra23}, the synchronization timescale is smaller by a factor $\sim 10$ than the prescription used in this work. As shown in the lower panel of Fig. \ref{tides_method}, one of the reasons is that the convective turnover timescales obtained with Eq. \eqref{pos} do not match the values predicted by the MLT approach. Two convective zones are present throughout the evolution of the models presented in Fig. \ref{tides_method}, consistently with the findings of e.g., \citet{mae08,he25}. The first one, the closest to the surface and the smallest in size and mass is due to the partial helium ionization zone, the second one, deeper in the envelope and larger in size, to the partial iron ionization zone. The helium zone being of very small size, Eq. \eqref{pos} predicts it to have $\tau_{\rm conv}\sim 10^2-10^3$\,s, whereas the MLT approach predicts $\tau_{\rm conv}\sim 5\cdot 10^6$\,s throughout the evolution. The mass and radius extent of the zone being very small, it does not contribute much to the total torque (see Eqs. \eqref{posydon} and \eqref{prescription}). The equilibrium tides are dominated by the other zone for which the two timescales differ by about one order of magnitude. 

Given that several terms differ between the present prescription (Eq. \eqref{prescription}) and that \citep[][i.e, Eq. \eqref{posydon}]{fra23}, it is not trivial to understand why the resulting synchronization timescales differ by one order of magnitude in Fig. \ref{tides_method}. The ratio of the \citet{fra23} timescale ($\tau_{\rm sync,F23}$) to the one obtained with Eq. \eqref{prescription} ($\tau_{\rm sync,S25}$) is:
\begin{equation}
\begin{split}
    \frac{\tau_{\rm sync,F23}}{\tau_{\rm sync,S25}}&=\frac{\left(\frac{k_2}{T}\right)_{\rm S25}}{\left(\frac{k_2}{T}\right)_{\rm F23}}\\&=\underbrace{\frac{21}{2}}_{\sim10}\underbrace{k_2}_{\sim10^{-2}}\underbrace{\frac{\tau_{\rm conv,F23}}{\tau_{\rm conv,MLT}}}_{\sim10^{-1}}\underbrace{\frac{f_{\rm conv,MLT}}{f_{\rm conv,F23}}}_{=1\ \text{close to sync.}}\underbrace{\frac{\frac{I_{\rm conv.reg.}}{I}}{\frac{M_{\rm conv.reg.}}{M}}}_{\sim10}\\&\sim 10^{-1}.
    \end{split}
    \label{order_mag}
\end{equation}
Eq. \eqref{order_mag} only gives an order of magnitude estimate for the system considered in Fig. \ref{tides_method}. The values of the different terms in the equations were empirically obtained from the models. Deviations between the two prescriptions larger than a factor 10 may arise depending on the initial parameters of the models. Eq. \eqref{order_mag} aims at illustrating that the two prescriptions contain factors divergent by up to several order of magnitudes. As a result the predicted synchronization timescales may well vary by more than an order of magnitude. For instance, over a full MS evolution, $k_2$ may vary in the range $10^{-2}-10^{-4}$ \citep[e.g.,][]{cla04,ros20a,ros22a,sci25}. Although in our simulations the mass of the convective shells are quite stable over the evolution (they slightly increase), their moment of inertia generally increase more due to the radial expansion of the star and the $r^2$ dependence of $I_{\rm conv.reg.}$, thus the last term in Eq. \eqref{order_mag} may vary accordingly.
\section{Evolution towards synchronization of the $P=1.4$\,days models of Sect. \ref{spin_down}}\label{AppC}
\begin{figure*}[b]
\begin{subfigure}[b]{0.5\textwidth}
\centering
\centerline{\includegraphics[trim=.32cm .3cm 0.6cm .35cm, clip=true, width=1.03\columnwidth,angle=0]{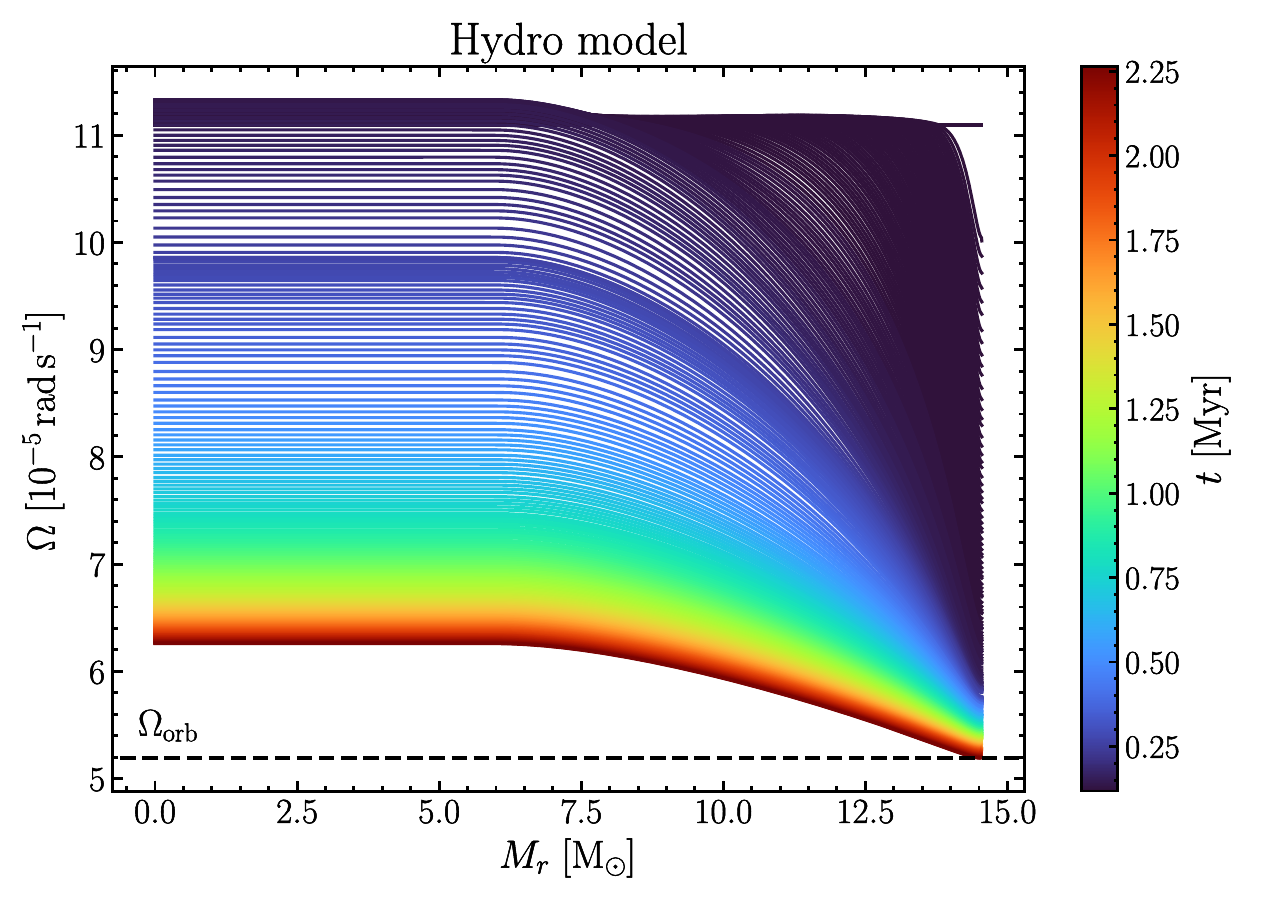}}
\end{subfigure}
\hspace{7pt}
\begin{subfigure}[b]{0.5\textwidth}
\centering
\centerline{\includegraphics[trim=.32cm .3cm 0.6cm .35cm, clip=true, width=1.03\columnwidth,angle=0]{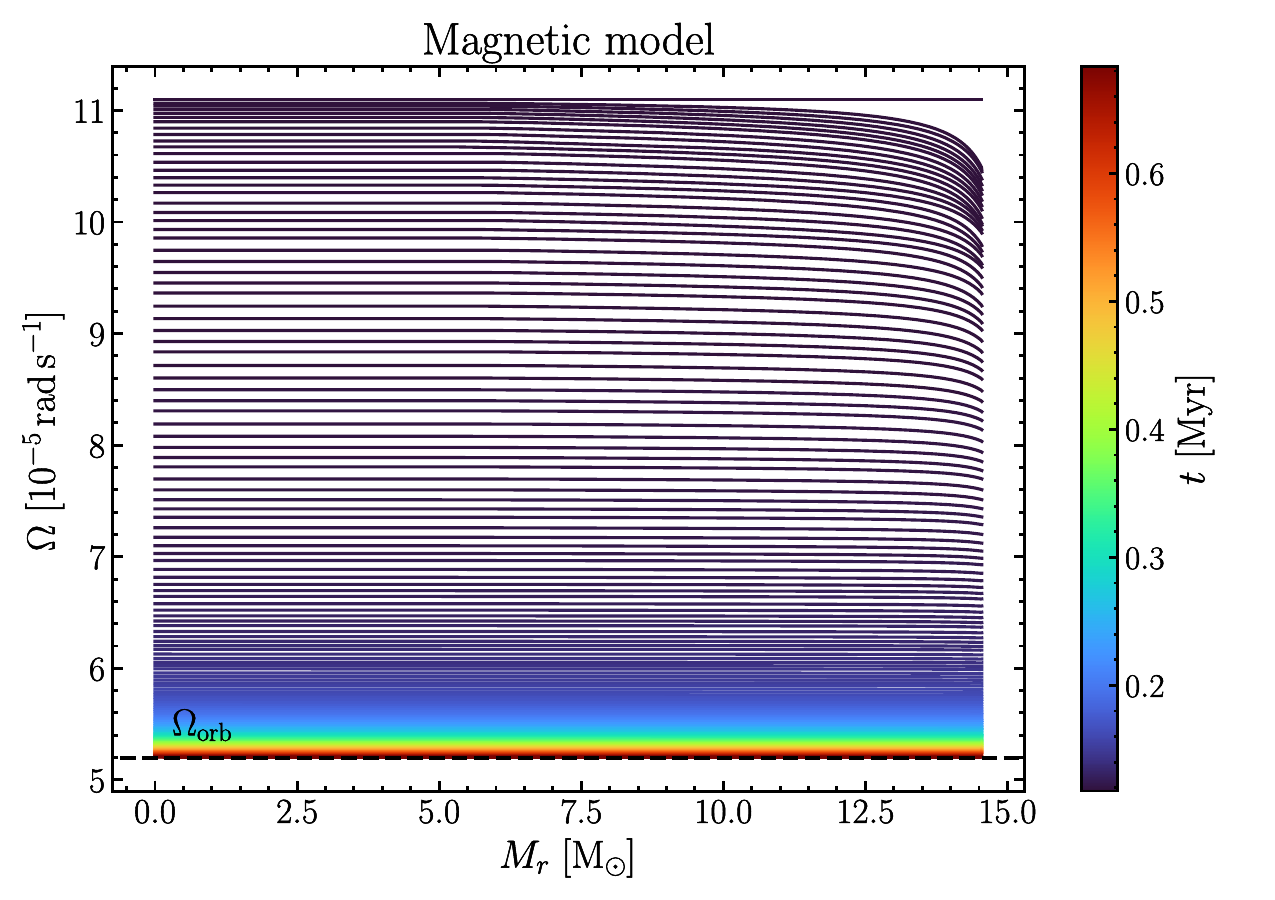}}
\end{subfigure}
\caption{$\Omega$--profiles of the $P=1.4$\,days models of Sect. \ref{spin_down} as they evolve towards synchronization. Each line represents a different epoch, indicated by its color. Dashed line: orbital angular velocity. \textit{Left panel:} hydro models. \textit{Right panel:} magnetic models.}
\label{omega_profiles}
\end{figure*}

Fig. \ref{omega_profiles} shows the $\Omega$--profiles of the models at $P = 1.4$\,days of Sect. \ref{spin_down}, as they evolve towards synchronization. In the hydro model, the tidal torque efficiently slows down the layers close to the surface. After $\sim 0.2$\,Myrs, they are already close to synchronization. In contrast, at the same age the core is still rotating super-synchronously, due to the mild core-envelope coupling. This corresponds to the moment when shear mixing is the most efficient. Interestingly, during the early evolution, the core slightly accelerates instead of slowing down, which can only be explained by the advective nature of the meridional currents. The surface being braked, diffusive processes can only transport AM outwards. Shortly after, the differential rotation becomes so important that AMT switches direction, slowing down the core. During this period, the surface layers are already close to synchronization, thus the tidal torque is not efficient anymore, (see Eq. \ref{correct} and remark below). The decay of the tidal torque, along with the AMT from core to surface delays the synchronization of the surface, which is achieved after $\sim2.25$\,Myrs.

In the magnetic model, the AMT is very efficient and the star rotates almost rigidly throughout the whole evolution. A slight degree of differential rotation is observed at the beginning of the evolution, when the tidal torque is maximal. We note that the star reaches synchronization after $\sim 0.7$\,Myrs, i.e., faster than the hydro model, which can seem counterintuitive. One might expect that the stronger the AMT, the more AM needs to be transferred to the star to reach synchronization, which should delay the moment when it is achieved. This picture is reasonably valid during the early evolution. After 5\,kyrs the surface velocity of the hydro model is lower than that of the magnetic model ($\Omega_{\rm spin,hydro}\sim 6.8\times 10^{-5}$\,rad\,s$^{-1}$, ${\Omega_{\rm spin,magn} \sim 7.5\times 10^{-5}}$\,rad\,s$^{-1}$). The dynamical tides formalism by \citet{zah77} predicts that the tidal torque scales as ${(\Omega_{\rm orb}-\Omega_{\rm spin})^{8/3}}$, the synchronization timescale as $|\Omega_{\rm orb}-\Omega_{\rm spin}|^{-5/3}$. The tidal torque gets extremely inefficient close to synchronization (the synchronization timescale diverges), and in the hydro model, unlike the magnetic model, the core is still rotating super-synchronously when this occurs. As a result, it experiences a long phase where AMT is transported from the core to the surface, delaying synchronization.

\section{Profiles of the hydro models of section \ref{song_massive}}\label{AppD}
\begin{figure*}[h]
\centering
\centerline{\includegraphics[trim=3.15cm 2.2cm .35cm 2.22cm, clip=true, width=1.4\columnwidth,angle=0]{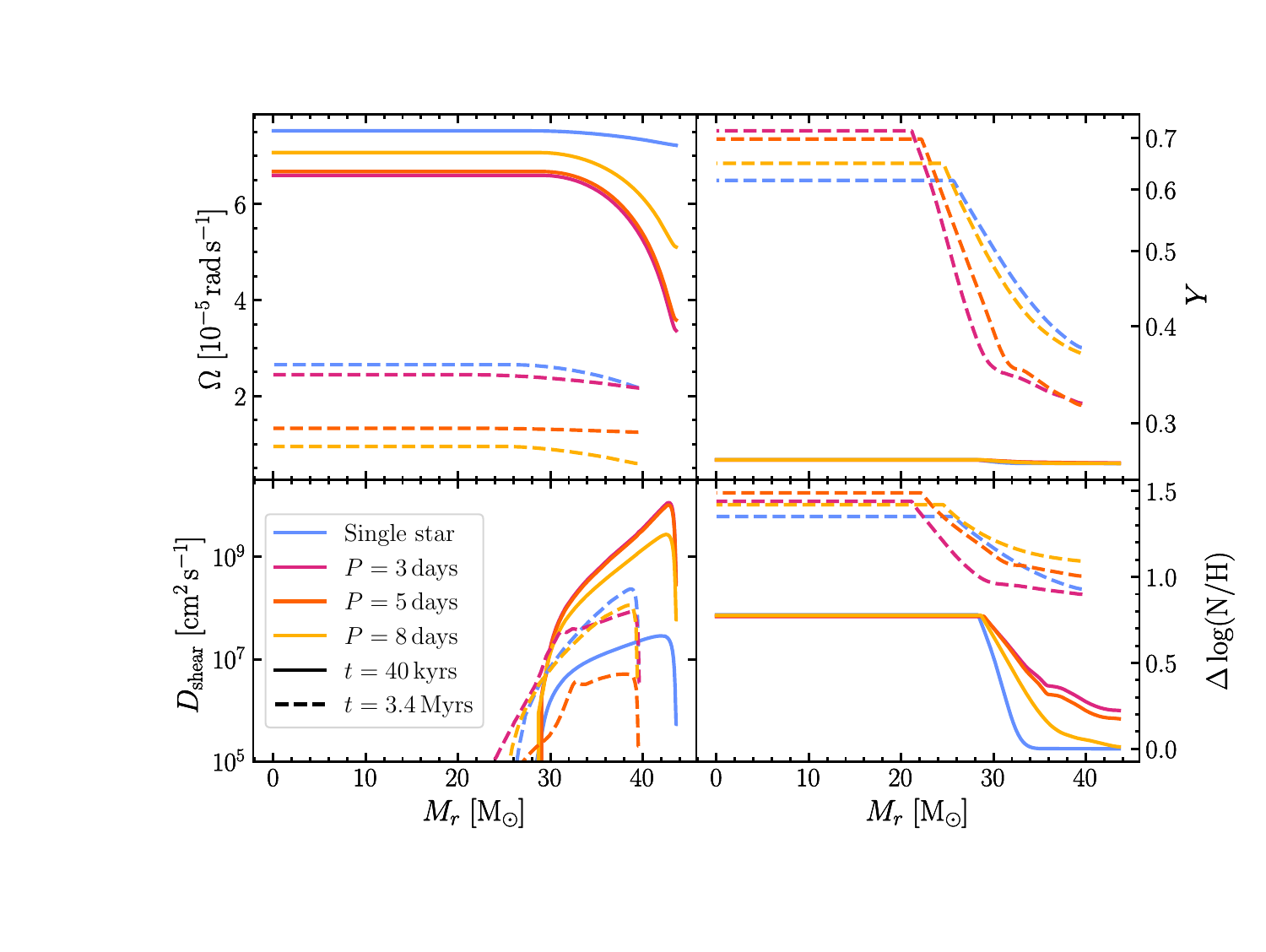}}
\caption{Same as Fig. \ref{profiles} but for the hydro models of Sect. \ref{song_massive}. Additionally, the helium mass fraction and nitrogen enhancement profiles are displayed. The profiles are shown at $t=40$\,kyrs and $t=3.4$\,Myrs.}
\label{profiles_massive}
\end{figure*}
The angular velocity, shear diffusion coefficient, helium mass fraction and nitrogen enhancement profiles of the hydro models of section \ref{song_massive} are displayed in Fig. \ref{profiles_massive} at $t=40$\,kyrs and $t=3.4$\,Myrs. At $t=40$\,kyrs the binary models are evolving towards synchronization and therefore have sharp $\Omega$-gradients. As a result, they have much larger shear coefficients than the single-star model. The nitrogen profiles are already significantly altered by CNO burning, and as a result, RM is efficient in diffusing nitrogen. Hence, the models show surface nitrogen enhancements. In contrast, the helium profiles are still almost flat, which is why RM is not efficient in enhancing helium at the surface at this point of the evolution, even though the shear coefficients are large. Later in the evolution, at $t=3.4$\,Myrs (which corresponds to the moment when the primary of the system with $P=3$\,days overfills its RL), the binary models have been synchronized for a long time and rotate with lower angular velocities and $\Omega$-gradients than the single-star model. As a result, the single-star model has the highest shear coefficient. At this point, the model already shows larger helium enrichment than the binary models, and a larger nitrogen enrichment than the model with $P=3$\,days. Later in the evolution and before the binary models with $P=5$ and 8\,days fill their RL (see Fig. \ref{spin_down_massive}), even its nitrogen enhancement gets larger than those of the binary models.
\section{Differential rotation in the hydro models of Sect. \ref{grid}}\label{AppE}
\begin{figure}[h]
\centering
\centerline{\includegraphics[trim=.cm .2cm .2cm .3cm, clip=true, width=1\columnwidth,angle=0]{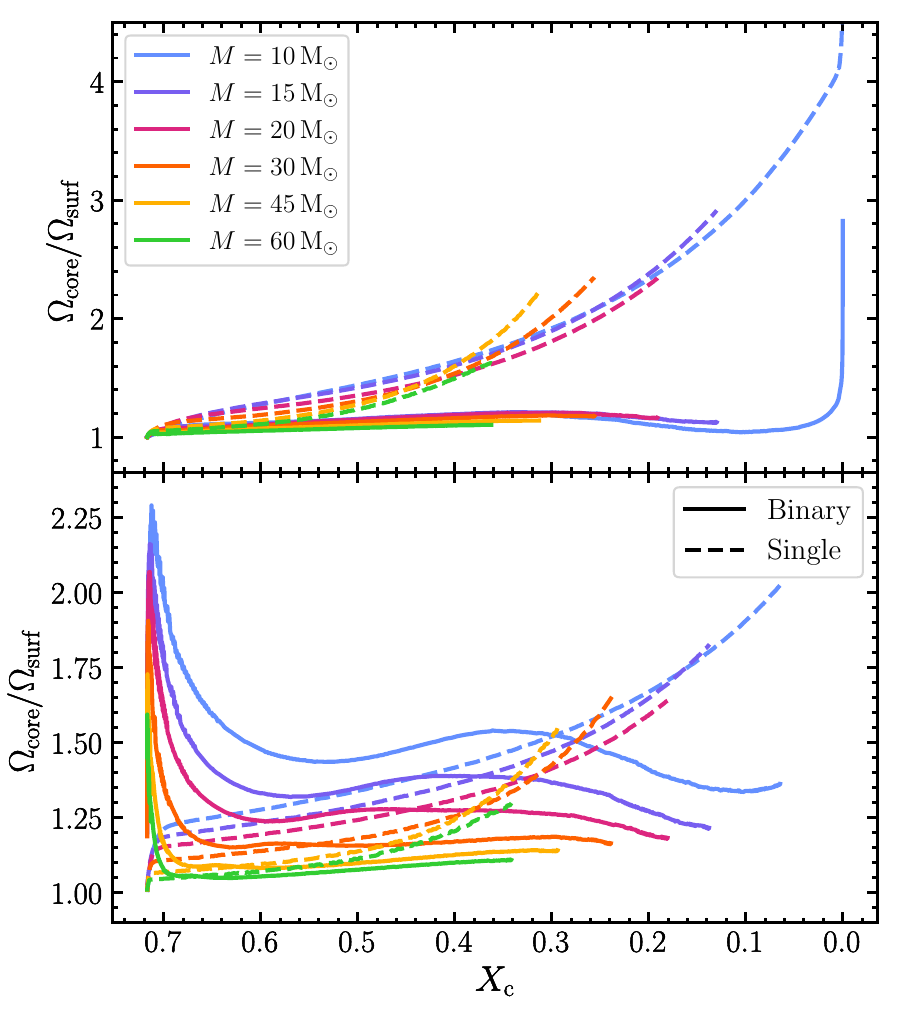}}
\caption{Time evolution of the core-to-surface rotation ratio of the hydro models of the grids of Sect. \ref{grid} at $P=3$\,days. \textit{Upper panel:} first grid (models initialized at synchronization). \textit{Lower panel:} second grid (models initialized at $\upsilon_{\rm ini}/\upsilon_{\rm crit}=0.4$).}
\label{rapom}
\end{figure}

Fig. \ref{rapom} shows the evolution of the core-to-surface rotation ratio of the hydro models of the two grids of Sect. \ref{grid} at period $P=3$\,days. The single-star evolutions are shown as a comparison. In the whole mass range, binary models initialized at synchronization evolve with a lower core-to-surface rotation ratio than the single stars initialized with identical velocities. Depending on the initial mass, in single-star models $\Omega_{\rm core}/\Omega_{\rm surf}$ reaches values from $\sim 1.6$ (60\,M$_\odot$) to $\sim 4$ (10\,M$_\odot$). Smaller degrees of differential rotation are reached in higher mass models because the AMT is more efficient and the simulations are stopped earlier\footnote{The binary models are stopped at RLOF or at TAMS, the single-star models are stopped when their central hydrogen mass fraction equates that of the last computed binary model.}.

In contrast, in binary models $\Omega_{\rm core}/\Omega_{\rm surf}$ does not exceed\;$\sim 1.2$ (the fast increase at the end of the evolution in the 10\,M$_\odot$ model is due to its end-MS contraction), which is why tidal torques may suppress rotational mixing in this case. By maintaining synchronization, tidal interactions reduce the $\Omega$--gradients, thereby diminishing the efficiency of shear mixing.

Binary models of the second grid reach high degrees of differential rotation ($\Omega_{\rm core}/\Omega_{\rm surf}\sim 2.3$ for the 10\,M$_\odot$ model, $\Omega_{\rm core}/\Omega_{\rm surf}\sim 1.6$ for the 60\,M$_\odot$ model) in the early evolution, due to the braking of the surface  by tidal torques. Subsequently, the core-to-surface rotation ratio decreases as AM is transported from the core to the surface. The AMT is more efficient at higher mass, which is the reason why $\Omega_{\rm core}/\Omega_{\rm surf}$ drops faster for these models. At the end of the evolution, models have $\Omega_{\rm core}/\Omega_{\rm surf} \sim 1.1-1.2$, consistently with those of the first grid. In single-star models, the degree of differential rotation steadily increases along the MS evolution, reaching values up to $\sim 2$. This illustrates that in the hydro binary models of the second grid, the spin-down by tides boosts rotational mixing in the early evolution, by increasing the $\Omega$--gradients.
\end{appendix}
\end{document}